\documentclass[fleqn,usenatbib]{mnras}

\usepackage{newtxtext,newtxmath}

\usepackage[T1]{fontenc}

\DeclareRobustCommand{\VAN}[3]{#2}
\let\VANthebibliography\thebibliography
\def\thebibliography{\DeclareRobustCommand{\VAN}[3]{##3}\VANthebibliography}

\usepackage{graphicx}	
\usepackage{amsmath}	
\defcitealias{arnaudova2025}{A25}
\defcitealias{drake2024lofar}{Dr24}
\usepackage{booktabs}
\newcommand{\HL}{\hline}

\title[The DESI View of LoTSS DR2]{The DESI View of the Faint Radio Source Population in LoTSS DR2}

\author[Arnaudova et al.]{M. I. Arnaudova$^{1,2}$\thanks{E-mail: m.i.arnaudova@gmail.com},
L. R. Holden$^{2}$,
D. J. B. Smith$^{2}$,
P. N. Best$^{1}$, 
R. Kondapally$^{3,4}$, 
K. J. Duncan$^{1}$,
\and
A. Bushi$^{1}$,
S. Das$^{2}$,
S. R. Flury$^{1}$,
C. L. Hale$^{1}$,
M. J. Hardcastle$^{2}$,
H. J. A. R$\ddot{\rm{o}}$ttgering$^{5}$
and S. Shenoy$^{2}$
\\
$^{1}$Institute for Astronomy, University of Edinburgh, Royal Observatory, Blackford Hill, Edinburgh, EH9 3HJ, UK\\
$^{2}$Centre for Astrophysics Research, University of Hertfordshire, Hatfield, AL10 9AB, UK\\
$^{3}$Centre for Extragalactic Astronomy, Department of Physics, Durham University, Durham DH1 3LE, UK\\
$^{4}$Institute for Computational Cosmology, Department of Physics, University of Durham, South Road, Durham DH1 3LE, UK\\
$^{5}$Leiden Observatory, Leiden University, PO Box 9513, NL-2300 RA Leiden, The Netherlands\\
}

\date{Accepted 2026 July 6. Received 2026 July 2; in original form 2026 February 23}

\pubyear{\the\year{}}

\begin{document}
\label{firstpage}
\pagerange{\pageref{firstpage}--\pageref{lastpage}}
\maketitle

\begin{abstract}
\noindent The faint radio-source population includes galaxies dominated by both star formation (SF) and active galactic nuclei (AGN), which are two key processes shaping galaxy evolution. To investigate this population, we probabilistically classified 251,413 radio sources from the second data release of the LOFAR Two-metre Sky Survey (LoTSS DR2) using spectroscopic data from the first release of the Dark Energy Spectroscopic Instrument (DESI DR1). Our classification method includes: (i) the identification of radio excess relative to SF, (ii) the Baldwin, Philips \& Terlevich (BPT) diagram, (iii) a modified Mass Excitation ($\mathcal{M}$Ex) diagram, and (iv) the [\textsc{Oiii}]$\lambda5007$ equivalent width. These are combined with Monte Carlo methods to estimate the probability that each source is a star-forming galaxy (SFG), a radio-quiet AGN (RQ AGN), or a low- or high-excitation radio galaxy (LERG or HERG), allowing various thresholds to be applied depending on science goals. Considering classifications above a 90 per cent probability threshold, we identify 68,820 SFGs, 32,288 RQ AGN, 35,210 LERGs and 3,085 HERGs, representing the largest radio sample to date with high-confidence spectroscopic classifications. Using this sample, we show with higher statistical power than previous studies that LERGs typically accrete below 1 per cent of the Eddington limit and HERGs above it. We also identify a small subset of high-accreting LERGs whose stacked spectra reveal prominent star-forming features, highlighting difficulties in interpreting their accretion properties. Our results demonstrate the power of large spectroscopic samples to characterise the radio-source population, providing a foundation for studies in the SKA era.

\end{abstract}




\section{Introduction}

Radio observations provide a dust-unbiased view of the energetic processes that shape galaxy evolution — from supernova-driven synchrotron and thermal (free-free) emission in star-forming galaxies (SFGs; \citealt{condon1992radio}) to the relativistic jets and outflows associated with active galactic nuclei (AGN; \citealt{begelman1984theory}) — making radio surveys a key tracer of cosmic activity. Investigating the radio source population is therefore essential for understanding how galaxies and their supermassive black holes have co-evolved over cosmic time. However, disentangling the dominant source of radio emission in each source is a non-trivial task and requires additional information (e.g. \citealt{williams2019}; \citealt{kondapally2021lofar}; \citealt{hardcastle2023}).

Based on the dominant source of emission, radio sources can be divided into four classes: star-forming galaxies (SFGs), whose emission is associated with stellar processes; radio-quiet AGN (RQ AGN), whose radio emission may be powered by star formation or by weak AGN-related processes such as small-scale jets or AGN-driven winds (e.g. \citealt{panessa2019}; \citealt{macfarlane2021radio}; \citealt{morabito2022identifying}; \citealt{petley2022}; \citealt{yue2024} \citealt{white2025}); and radio-loud AGN, which possess powerful radio jets and can be further divided into high-excitation radio galaxies (HERGs) and low-excitation radio galaxies (LERGs), depending on their accretion properties \citep[for a review, see e.g. ][]{heckman2014coevolution, hardcastle2020radio}. In this scheme, radio-quiet AGN and HERGs are radiatively-efficient systems that are believed to accrete material at relatively high rates (typically above 1 per cent of the Eddington limit, $\lambda_{\mathrm{Edd}} \gtrsim 0.01$) through a geometrically thin, optically thick accretion disc (e.g. \citealt{shakura1973}), in which matter is efficiently converted into radiation across the electromagnetic spectrum. In contrast, LERGs are radiatively-inefficient AGN thought to accrete at lower accretion rates ($\lambda_{\mathrm{Edd}} \lesssim 0.01$) via a geometrically thick, optically thin flow (\citealt{narayan1994, narayan1995}), which does not efficiently convert matter into radiation. As a result, they show little or no radiative AGN signatures at other wavelengths but appear highly efficient at producing relativistic radio jets.

In recent years, distinguishing between these populations observationally has become increasingly feasible thanks to the ability to cross-match radio surveys with multi-wavelength imaging and spectroscopic surveys (e.g. \citealt{best2005}; \citealt{best2012on}; \citealt{gurkan2014}; \citealt{pracy2016}; \citealt{smolcic2017}; \citealt{whittam2022}; \citealt{best2023lofar}; \citealt{das2024lofar}; \citealt{drake2024lofar}; \citealt{hardcastle2025}; \citealt{arnaudova2025}). New-generation radio interferometers, such as the Low-Frequency Array (LOFAR; \citealt{vanhaarlem2013lofar}), have also hugely increased the sample size of radio sources that can be studied. In particular, the LOFAR Two-metre Sky Survey (LoTSS) observes the faint radio sky at 144~MHz with high sensitivity and resolution, enabling the identification of large samples of powerful radio sources across a wide area (LoTSS-wide; \citealt{shimwell2017, shimwell2019, shimwell2022lofar, shimwell2026}), as well as fainter populations in deeper fields (LoTSS Deep fields; \citealt{tasse2021lofar}; \citealt{sabater2021lofar}; \citealt{shimwell2025}). 

The LoTSS Deep fields observations allowed \citet{best2023lofar} to employ multi-wavelength photometry and four independent spectral energy distribution (SED) fitting codes (\texttt{\textsc{MAGPHYS}}; \citealt{daCunha2015}, \texttt{\textsc{BAGPIPES}}; \citealt{carnall2018}, \texttt{\textsc{CIGALE}}; \citealt{noll2009}; \citealt{boquien2019} and \texttt{\textsc{AGNFITTER}}; \citealt{calistro2016}) to produce maximum likelihood classifications of $\approx$ 80,000 sources in the LoTSS Deep Fields Data Release 1. Building upon this, \citet{das2024lofar} showed that similar results could be achieved using a single SED-fitting code (\texttt{\textsc{PROSPECTOR}}; \citealt{leja2018}; \citealt{johnson2021}), significantly simplifying the framework and reducing the computational resources required. However, these photometric classifications rely on model assumptions that are subject to degeneracies between stellar age, dust content, and metallicity, and often on photometric redshift estimates, which can limit the accurate identification of AGN, particularly when the galaxy’s overall SED is dominated by star formation. Using spectroscopic information from the Dark Energy Spectroscopic Instrument (DESI; \citealt{desi2026a, desi2016b}) for a sub-sample of sources photometrically classified by \citet{best2023lofar} and \citet{das2024lofar}, \citet[][hereafter \citetalias{arnaudova2025}]{arnaudova2025} demonstrated that the two photometric classifications miss a substantial fraction of radiatively efficient AGN. Specifically, they found that the photometric approach underestimates the number of RQ~AGN by a factor of $\sim$3-5 compared to spectroscopic identifications at >90 per cent confidence. Furthermore, when studying the Eddington-scaled accretion rates of LERGs and HERGs, \citetalias{arnaudova2025} showed that photometric classifications obscure the distinction between the two accretion modes, whereas spectroscopic diagnostics recover two clearly separated distributions, as had been suggested by e.g. \cite{best2012on} using bright radio samples at low redshifts ($z<0.3$). Spectroscopy therefore remains the preferred method for distinguishing between these classes and can be applied in wider fields with limited multi-wavelength data, such as the LoTSS-wide area, which contains a substantially larger number of sources than the deep fields. The LoTSS second data release (LoTSS DR2) is currently the largest low-frequency radio catalogue with optical counterparts, comprising over 4.1 million sources across 5700~deg$^2$ (see \citealt{hardcastle2023} for details).

\citet[][hereafter \citetalias{drake2024lofar}]{drake2024lofar} presented a probabilistic spectroscopic classification scheme for the subset of LoTSS~DR2 sources with spectra from the Sloan Digital Sky Survey (SDSS), classifying over 150,000 sources up to $z\sim0.6$. This framework combined a radio-excess diagnostic, based on the relation between the dust-corrected H$\alpha$ and radio luminosity, with the BPT-NII diagram (\citealt{baldwin1981}), and employed Monte Carlo methods to determine the probability of each source being an SFG, RQ~AGN, LERG, or HERG. The authors validated their method by comparing their results with independent emission-line methods (e.g. the Excitation Index; \citealt{best2012on}), WISE selection techniques (e.g. \citealt{gurkan2014}), and physical properties found in the literature (e.g. \citealt{gurkan2018}; \citealt{smith2021lofar}; \citealt{das2024lofar}).

With the advent of DESI, this approach can be extended to a larger and deeper spectroscopic sample, probing a radio source population approximately twice as optically faint as that reached by SDSS. Moreover, by adopting the method introduced by \citetalias{arnaudova2025}, in which a modified Mass–Excitation ($\mathcal{M}$Ex) diagram is used in place of the BPT at higher redshifts, we are able to extend this classification scheme up to $z\sim1$. Furthermore, including the equivalent width of [\textsc{Oiii}]$\lambda5007$ diagnostic (e.g. \citealt{laing1994}; \citealt{tadhunter1998}; \citealt{best2012on}) allows classification of many sources left unclassified by \citetalias{drake2024lofar}. The significantly larger statistical sample compared to \citetalias{arnaudova2025}, together with the broader area covered, which enables the inclusion of rarer, more luminous jet populations, provides the opportunity to build upon their study of LERGs and HERGs and to further investigate their accretion-rate distributions.

This paper is structured as follows. Section~\ref{sec:data} describes the LoTSS and DESI datasets and our sample selection. Section~\ref{sec:fitting} details the spectral fitting and emission-line measurements. Section~\ref{sec:classifications} presents the probabilistic classification framework. Section~\ref{sec:results} presents a comparison with the SDSS classification scheme from \citetalias{drake2024lofar}, and discusses the selection effects introduced by using DESI spectroscopy. Section~\ref{sec:properties_lergs_hergs} focusses on the Eddington-scaled accretion rates of the radio-loud population, while section~\ref{sec:summary} summarizes our results. Throughout, we adopt a flat $\Lambda$CDM cosmology with $H_{0} = 70~\mathrm{km~s^{-1}~Mpc^{-1}}$, $\Omega_{\mathrm{M}} = 0.3$, and $\Omega_{\Lambda} = 0.7$.  

\section{Data}\label{sec:data}

\subsection{LoTSS DR2-wide}\label{sec:radio data}

The radio data used in this work are taken from LoTSS DR2 (\citealt{shimwell2022lofar}). This dataset consists of observations centred at 144\,MHz with a resolution of 6 arcsec (using the Dutch stations only), covering $\sim5700$ deg$^2$ of the northern sky (see the black outline of the footprint in the left panel of Figure \ref{fig:moc}) with a median rms sensitivity of 83\,$\mu$Jy beam$^{-1}$. Using this data release, \cite{hardcastle2023} employed the likelihood ratio method (\citealt{sutherland1992likelihood}; \citealt{smith2011herschel}), complemented by visual classifications from the LOFAR Galaxy Zoo\footnote{\href{http://lofargalaxyzoo.nl/}{http://lofargalaxyzoo.nl/}} to account for extended and complex radio morphologies (following the approach of \citealt{williams2019}), providing reliable optical identifications for 3.5 million sources (about 85\% of the total sample). Approximately 58\% of the LoTSS DR2 sources have robust redshift estimates, including photometric redshifts from \cite{duncan2022}, as well as spectroscopic redshifts from SDSS DR16 (\citealt{ahumada2020}) and DESI DR1 (\citealt{desi2025}), and, as a result, they serve as the starting point for our classification sample.

\subsection{DESI DR1}\label{sec:desi data}

\begin{figure*}
    \centering
    \includegraphics[width=1\textwidth]{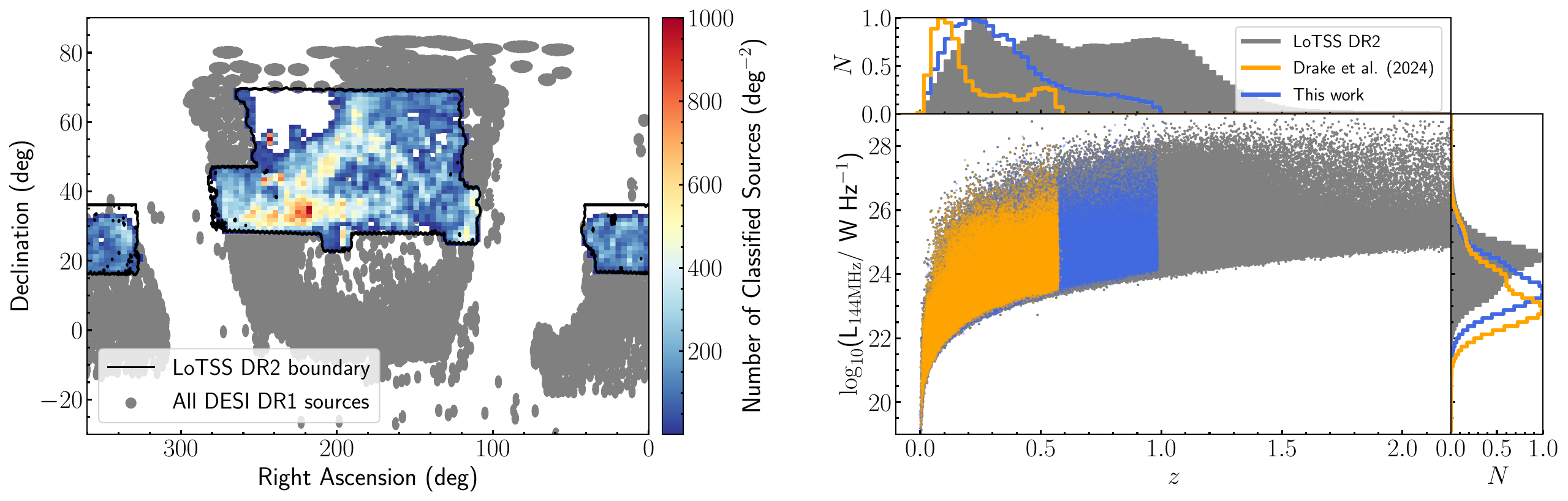}
    \caption{Left panel: Sky coverage of DESI DR1 and LoTSS DR2. The grey region indicates all DESI DR1 sources, the black lines show the LoTSS DR2 footprint, while the colour scale represents the number density of sources in our classification sample. Right panel: The redshift-radio rest-frame luminosity plane for all radio sources with a multi-wavelength counterpart (in grey), the sources with spectroscopic information from DESI DR1 (blue), and those that are part of the SDSS sample from \citetalias{drake2024lofar} (orange), along with their respective luminosity and redshift distributions (normalised to the maximum bin count in each sample). For reference the total number of sources ($N$) in each sample is $N_{\rm LoTSS} = 4,167,359$ (of which 2,392,639 have a photometric or spectroscopic redshift), $N_{\rm DESI}=279,700$, and $N_{\rm SDSS}=152,355$. }
    \label{fig:moc}
\end{figure*}

The spectroscopic data used in this work are taken from DESI DR1 (\citealt{desi2025}). This dataset is currently the largest extragalactic dataset with spectroscopic redshifts, consisting of a total of 18.7 million objects across 14,000 deg$^{2}$ area (shown in grey in Figure \ref{fig:moc}). The targeting algorithm for these objects is associated with optical and mid-infrared photometry from the Legacy Imaging Surveys (\citealt{dey2019overview}) and the Wide-field Infrared Survey Explorer (WISE; \citealt{wright2010}), respectively, and is broadly divided into the following categories: the Milky Way survey (MWS; \citealt{cooper2023}), Bright Galaxy Survey (BGS) galaxies ($0 < z < 0.6$; \citealt{ruiz2020prelimenary}; \citealt{hahn2023desi}); Luminous Red Galaxies (LRGs, $0.4 < z < 1.1$; \citealt{zhou2020prelimenary, zhou2023target}); Emission-line Galaxies (ELGs, $0.6 < z < 1.6$; \citealt{raichoor2020prelimenary, raichoor2023target}); QSOs ($0.9 < z < 4$; \citealt{yeche2020prelimenary}; \citealt{Chaussidon2023target}), as well as a range of secondary targets (SCND; see e.g \citealt{myers2023} for details). The spectra were taken using 10 identical spectrographs mounted on the 4m Mayall Telescope at Kitt Peak National Observatory, each equipped with 500 fibres with a 1.5$\arcsec$ entrance diameter, covering the wavelength range of $3600 - 9824$\AA\ at a resolving power of $\Delta\lambda/\lambda\approx2000-5500$ (\citealt{guy2023spec}). Together with the LoTSS DR2 optical identifications, these high-quality DESI spectra allow us to construct a large sample of radio sources suitable for robust classification into SFGs, RQ AGN, LERGs and HERGs.

\subsection{Sample selection}\label{sec:sample}

Starting with the LoTSS DR2 optical identifications from \cite{hardcastle2023}, we perform a positional cross-match with the DESI DR1 catalogue using a maximum search radius of 1.5 arcseconds and the recommended quality flags. Specifically, we select unique spectra (ZCAT\_PRIMARY == True) corresponding to non-sky targets (OBJTYPE == "TGT") with reliable redshift measurements (ZWARN == 0). This results in a total of 366,122 radio sources with available spectroscopic information, of which 363,326 are extragalactic (SPEC\_TYPE != `STAR').
Next, we consider the aperture corrections provided by the \textsc{\texttt{FastSpecFit}} Spectral Synthesis and Emission-Line Catalogue (\citealt{Moustakas2023fastspecfit}; Moustakas et al. \textit{in prep.}), one of DESI's value added catalogues, which we use to correct flux losses in the emission-line measurements required for our classification (see section \ref{sec:classifications}). We note that \textsc{\texttt{FastSpecFit}} uses updated redshifts from \texttt{\textsc{QuasarNET}}\footnote{\href{https://github.com/ngbusca/QuasarNET}{https://github.com/ngbusca/QuasarNET}} for QSO targets, while the value added catalogue from \cite{Siudek2024}, which we later use for the classifications (see section \ref{sec:classifications}), and the redshifts used for the spectral fitting are estimated with \textsc{\texttt{redrock}}\footnote{\href{https://github.com/desihub/redrock}{https://github.com/desihub/redrock}}. Therefore, we remove 1,285 sources which have $\Delta z>0.001$. As in \citetalias{arnaudova2025}, we exclude sources requiring an aperture correction factor greater than 5, as well as sources having an aperture correction less than 1, resulting in a sample of 307,732 radio sources. While \citetalias{arnaudova2025} applied aperture corrections based on the total-to-fibre $r$-band flux ratio, we find that this correction systematically overestimates the H$\alpha$-derived SFRs (by a median offset of $\sim0.3$\,dex) relative to those in the MPA–JHU catalogue (\citealt{kauffmann2003mpa}; \citealt{Brinchmann2004}; \citealt{tremonti2004}). In contrast, the \texttt{\textsc{FastSpecFit}} aperture corrections yield much better agreement (with a median offset of only 0.06\,dex), and therefore we adopt them in this work.

Finally, since our classification scheme (see section \ref{sec:classifications}) relies on the availability of specific emission lines, we restrict our sample to $z<0.947$ (hereafter the parent sample) and consider the following redshift ranges:

\begin{itemize}
    \item $0<z<0.483$ (hereafter the low-$z$ sample), which includes 207,657 sources with spectral data covering the H$\alpha$ and [\textsc{Nii}]\,$\lambda$6583 emission lines, given the wavelength coverage of the DESI spectrographs, necessary for the use of the radio diagnostic and the BPT-NII diagram (\citealt{baldwin1981}).
    
    \item $0.483<z<0.947$ (hereafter the high-$z$ sample), which includes 72,043 sources with spectral data covering the H$\beta$ and [\textsc{Oiii}]\,$\lambda$5007 emission lines, necessary to use the modified Mass-Excitation diagram (hereafter $\mathcal{M}$Ex; \citetalias{arnaudova2025}).
\end{itemize}

The source density of these sources and their distribution in the redshift - radio rest-frame luminosity ($L_{144\mathrm{MHz}}$\footnote{The $L_{144\mathrm{MHz}}$ is calculated from the integrated 144 MHz flux density ($S_{\mathrm{144\,MHz}}$; \citealt{hardcastle2023}), assuming a spectral index of $\alpha = -0.7$ ($S_{\nu} \propto \nu^{\alpha}$). Redshifts are taken from DESI DR1 for the sample analysed in this work, from the \citetalias{drake2024lofar} catalogue for the SDSS sample, and from the spectroscopic and photometric redshifts compiled in \citet{hardcastle2023} for the full LoTSS sample.}) plane are presented in Figure \ref{fig:moc}. In the right panel, we can see that we are probing a parameter space which differs from that available to \citetalias{drake2024lofar}; due to DESI's selection criteria, which rely on the Legacy Surveys which reach approximately 1.5 mag deeper than SDSS in the $g$, $r$, and $z$ bands, we have spectroscopic measurements for about twice as many radio sources at $z<0.6$ (corresponding to the redshift range of \citetalias{drake2024lofar}). Furthermore, as a result of the updates to the classification scheme introduced in \citetalias{arnaudova2025}, we extend the redshift range studied up to $z\sim1$. 
We note that the impact of selection effects, both relative to \citetalias{drake2024lofar} and in general, is discussed in section \ref{sec:results}. While this parameter space still differs from that of the full LoTSS DR2 catalogue, the WEAVE-LOFAR survey (\citealt{smith2016}), which will select targets solely based on their 144\,MHz flux density, will provide a more uniform and complete coverage of the full radio-source population.

\section{Spectral Fitting}\label{sec:fitting}

For this work and in preparation for the WEAVE-LOFAR survey, we have developed a new spectral-fitting code, \texttt{\textsc{WL-SLAYER}}: the WEAVE-LOFAR Spectral Line Analyser (see \citealt{arnaudova2024b}, \citealt{ holdentadhunter2025}; \citetalias{arnaudova2025}, and \citealt{holden2025} for predecessors), in order to obtain robust emission line measurements and realistic uncertainties needed for the classification scheme (see section \ref{sec:classifications}). This new code is particularly necessary for radio sources, as it is designed to model complex emission line profiles that are commonly observed in these objects (e.g. due to AGN-driven winds or outflows; \citealt{molyneux2019extreme}; \citealt{Escott2025}). In this section, we give details of the approach and describe how we used it to produce line-flux measurements and various quality-assurance flags that we make use of in our analysis.

\subsection{Stellar continuum modelling and subtraction}\label{sec:fitting:preparation}

To isolate the flux from the emission lines, \texttt{\textsc{WL-SLAYER}} models and subtracts the stellar continuum using the \texttt{\textsc{pPXF}} code \citep{cappellari2004, cappellari2017, cappellari2023}. To do this, it first corrects the spectra for Galactic extinction using the $R_v=3.1$ \citet{fitzpatrick1999} extinction law (as implemented in the \textsc{Python} \textsc{\href{https://extinction.readthedocs.io/en/latest/}{extinction}} package) with the dust maps from \citet{schlegel1998}, and uses the redshift values provided in the DESI DR1 catalogue to shift the spectra to the rest-frame. Next, it masks regions spanning $\pm$750\;km\;s$^{-1}$ around common emission lines, ensuring that even broad Balmer lines do not affect the fit. Finally, it models the full wavelength range using a linear combination of \citet{bruzual2003} stellar templates (BC03), along with a fourth-order polynomial to correct low-frequency continuum variations, such as flux-calibration errors or dust extinction of the stellar continuum. The fits are performed in log-wavelength space, after which the resulting stellar-continuum model is interpolated onto a linear wavelength grid and subtracted from the spectrum, leaving only the emission-line flux for subsequent analysis. We note that this continuum fitting approach does not include AGN templates as the majority of the sample is expected to consist of star-forming galaxies and type 2 AGN whose continua are dominated by stellar emission. Type 1 AGN are identified and excluded from the final sample (see section \ref{sec:fitting:type1_flag}) and will be explored in future work (Bushi et al. \textit{in prep.}).

\subsection{Emission-line fitting}\label{sec:fitting:mcmc}

\begin{figure*}
    \centering
    \includegraphics[width=0.93\textwidth]{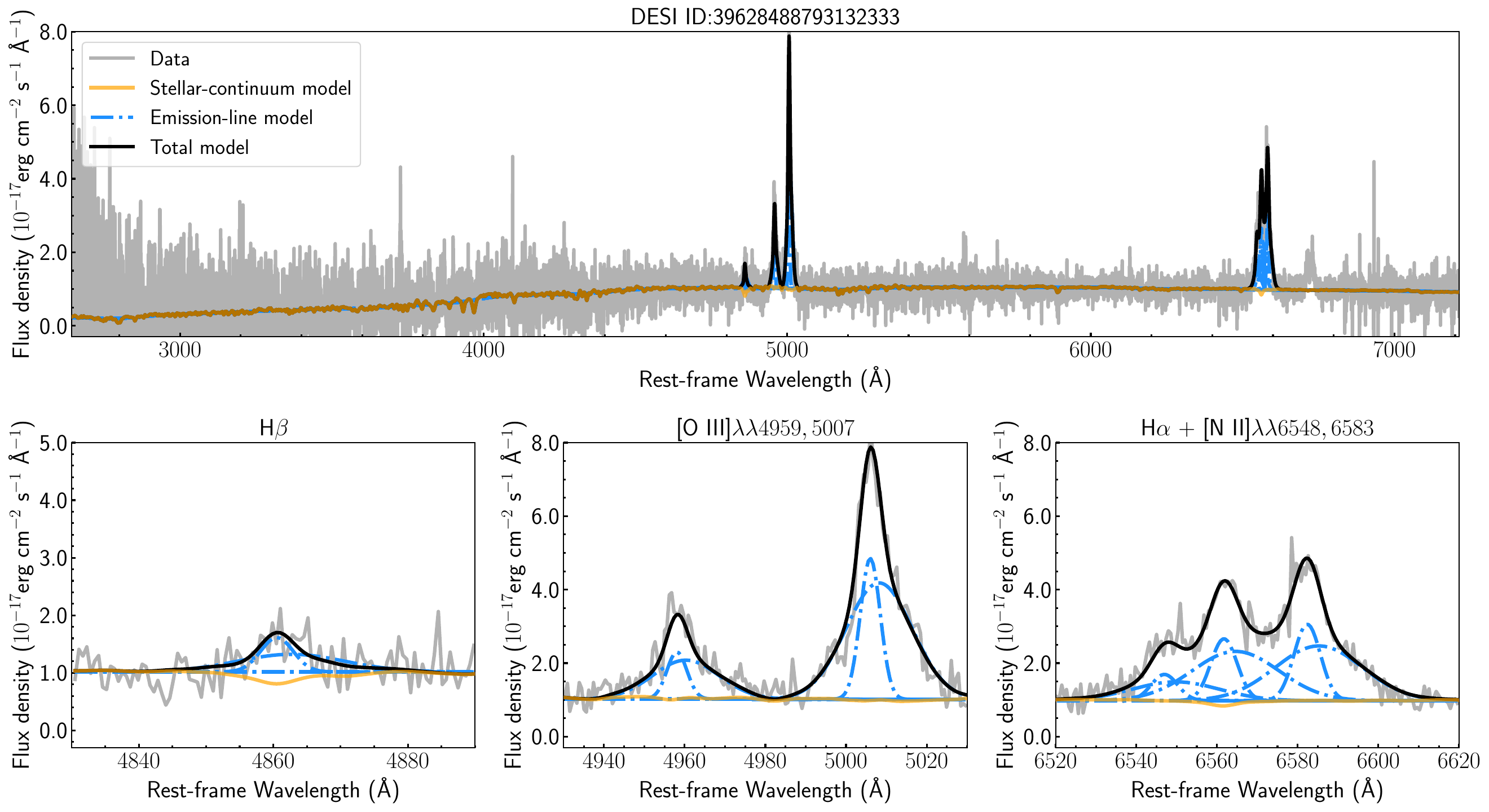}
    \caption{An example fit of a random galaxy spectrum using \texttt{\textsc{WL-SLAYER}}. The observed data are shown in grey, the continuum model in orange, the emission-line model in blue, and the total best-fitting model in black. The top panel displays the fit to the full spectrum, which is used for continuum subtraction, while the bottom panels focus on the H$\beta$, [\textsc{Oiii}]$\lambda\lambda4959,5007$, and H$\alpha$+[N\textsc{ii}] complexes to highlight the emission-line modelling.}
    \label{fig:spec_fit}
\end{figure*}

For the emission line modelling in this work, we focus only on the H$\beta$, $\mathrm{[O\,III]}\lambda\lambda4959,5007$, H$\alpha$, and $\mathrm{[N\,II]}\lambda\lambda6548,6583$ emission lines (for the high-$z$ sample we model only H$\beta$ and $\mathrm{[O\,III]}\lambda\lambda4959,5007$ as the rest fall outside the observing window). However, we note that \texttt{\textsc{WL-SLAYER}} is designed to work with any user defined list of lines (e.g. Holden et al. \textit{in prep.}). Here, we implement MCMC sampling using the Affine Invariant Ensemble provided by the \textsc{emcee Python} package \citep{foremanmackey2013}, with the Differential Evolution Markov Chain (DE-MC) algorithm for updating the walker positions \citep{terbraakvrugt2008}. The DE-MC is used to enable efficient sampling of the high-dimensionality parameter spaces that we are considering when fitting multiple emission line components simultaneously. Our emission-line model consists of a number of Gaussian components ($N_g$) for each emission line and is fitted in velocity space, with each Gaussian component having the same velocity shift and width for all lines but a variable peak flux (that is, if a broad component is required for the Balmer lines, the corresponding forbidden-line component would have an amplitude of zero).

To ensure that the MCMC walkers quickly converged to regions of maximum probability, our routine first performs a least-squares fit of the emission-line model using the \textsc{SciPy Python} library \citep{virtanen2020}, the results of which we use as initial positions for the walkers in parameter space (with a small scatter). Physically-motivated lower and upper bounds were placed on the parameters of the least-squares fit: the velocity widths of the Gaussian components were required to be positive, and constraints on the relative fluxes in line doublets were set based on atomic physics ($\mathrm{[O\,III]}(5007/4959) = 2.99$; $\mathrm{[N\,II]}(6583/6548) = 2.92$: \citealt{osterbrock2006}). The peak flux is allowed to be negative to account for the presence of absorption lines, as well as to avoid a `positive bias' effect in the presence of noise.
The same constraints were used as priors for the MCMC sampling itself, with the addition of Gaussian priors on velocity shift ($\mu=0$\;km\;s$^{-1}$; $\sigma=400$\;km\;s$^{-1}$) and width ($\mu=200$\;km\;s$^{-1}$; $\sigma=50$\;km\;s$^{-1}$) to ensure that walker positions were weighted against physically-unlikely extreme values. The values of the Gaussian priors were selected as a result of testing the fitting routine on DESI spectra presenting emission-lines of different types (e.g. weak lines, strong narrow lines, type 1 AGN), and result in walkers converging on reasonable velocity values (and hence avoiding anomalously large flux values arising from fitting very broad low-peak components) when there is little (or zero) emission line flux while still being able to converge on large velocity widths/shifts in the case of genuinely kinematically-extreme emission lines.

The log-likelihood function used in our MCMC routine was:
\begin{equation}
    \ln \mathcal{L}=-\frac{1}{2}{\sum^k_{i=1}}\Biggl(\frac{F_\mathrm{\lambda,i}-F^\mathrm{m}_\mathrm{\lambda,i}}{F^\mathrm{err}_\mathrm{\lambda,i}}\Biggl)^2,
    \label{eq:lnlike}
\end{equation}
where, at the velocity step $i$ (of maximum steps $k$), $F_\mathrm{\lambda,i}$ is the flux density at wavelength $\lambda$ in a given DESI spectrum, $F^\mathrm{m}_\mathrm{\lambda,i}$ is the modelled flux density, and ${F^\mathrm{err}_\mathrm{\lambda,i}}$ is the flux-density uncertainty for the DESI spectrum. In order to ensure that the MCMC walkers had converged on the region of maximum likelihood and therefore were effectively sampling the posterior distribution, we calculated the autocorrelation times ($\tau$) from the latter half of the walker chain for every parameter at 1000\;step intervals, and stopped sampling when all autocorrelation times were less than 5\;per\;cent of the current number of steps.

Another feature of \texttt{\textsc{WL-SLAYER}} is that it uses an iteratively-increasing number of Gaussian components to model emission lines. We chose the first iteration to use a single Gaussian component for each emission line ($N_g=1$): MCMC sampling for this model is performed as described earlier in this section. Once the walkers have converged, a fit with $N_g + 1$ components is performed, and the Bayesian Information Criterion (BIC) is calculated following:
\begin{equation}
	\mathrm{BIC} = k\mathrm{ln}(n) - 2\mathrm{ln}(\hat{L})
	\label{eq:bic}
\end{equation}
where $k$ is the number of parameters (dimensions) of the model of a given iteration, $n$ is the number of flux (or wavelength) data points from a given spectrum that is considered in the emission-line model, and $\mathrm{ln}\hat{L}$ is the maximised value of the log-likelihood function (Equation\;\ref{eq:lnlike}). If the decrease in the value of the BIC was greater than six (following the recommendation of \citealt{Raftery1995}), we accepted the more complex model and continued with further iterations until this criterion was not met. With this approach, we are able to simultaneously fit the profiles of different emission lines with a statistically justified number of Gaussian profiles, preventing under- and overfitting. 

An example output from this routine is shown in Figure \ref{fig:spec_fit}, where a randomly selected galaxy spectrum (solid grey line) is displayed alongside its best-fit model (solid black line). The corresponding stellar-continuum (solid orange line) and emission-line components (dashed-dotted blue line) are also presented.

\subsection{Emission-line parameter calculation}\label{sec:fitting:flux_calc}

In order to produce measurements of key emission-line parameters, including the total line flux, equivalent width (EW), and full-width-half-maximum (FWHM), we directly used the last 1000 steps of the flattened MCMC chains. For each of these steps, we generated a total line profile (consisting of the final number of Gaussians determined using the BIC) using the parameter values extracted from their respective flattened chains. The line flux for a given realisation was determined by calculating the total area of the line profile, which would be positive for a well-detected emission line, close to zero for a non-detection, and negative for an absorption line. We then used all the last 1,000 realisations for a given emission line to produce a line-flux distribution, from which we took the 50th percentile as the line-flux value and estimated $1\sigma$ uncertainties using the 16th and 84th percentiles. A similar procedure was done for the EW and FWHM measurements.

\subsection{Identifying Type 1 AGN}\label{sec:fitting:type1_flag}

In addition to the emission-line parameters for our catalogue, we have defined a quality-control flag that identifies likely Type\,1 AGN. This is useful for excluding sources in which a significant fraction of the recombination-line flux (e.g. H$\alpha$, H$\beta$) does not originate from the same region as the forbidden lines, leading to contamination of crucial flux ratios that are used for source classification.

\begin{figure}
    \centering
    \includegraphics[width=\linewidth]{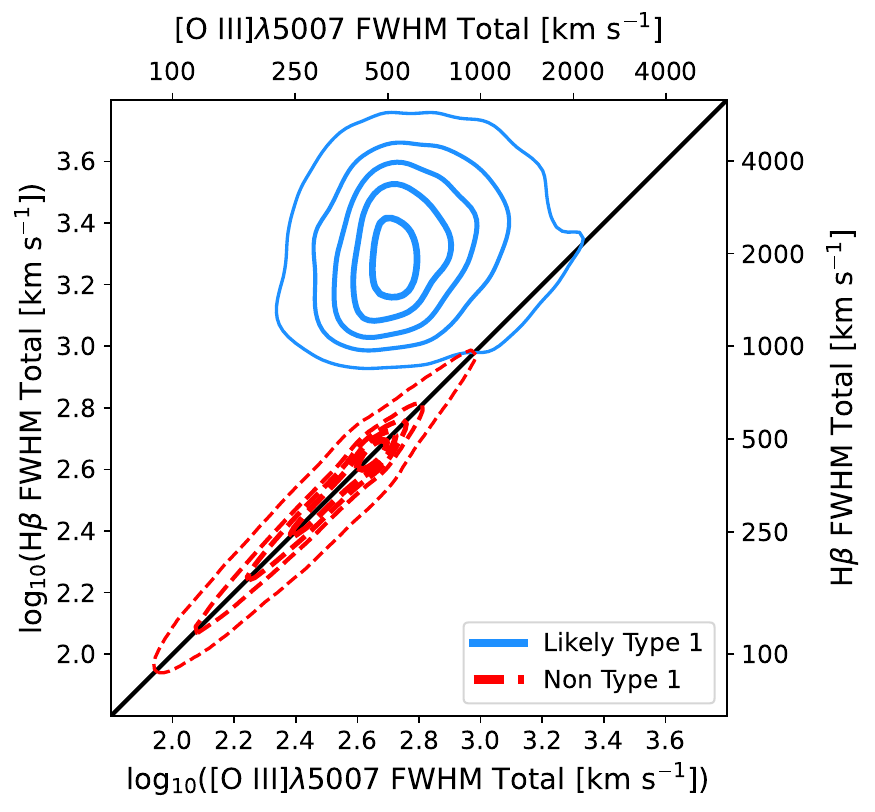}
    \caption{FWHM values for the total $\mathrm{[O\,III]}\lambda5007$ and H$\beta$ line profiles measured by \texttt{\textsc{WL-SLAYER}}. The probability density function for the objects in our sample that are identified as likely Type\,1 AGN by our \textsc{flag\_type1} flag are shown as solid blue contours, while objects not flagged as Type\,1 AGN are shown as dashed red contours.. Contours contain 10, 30, 50, 70, and 90\;per\;cent of the total objects for each flag value.}
    \label{fig:type1_vel_dist}
\end{figure}

For each realisation drawn from the MCMC chains, we measured the FWHM of the total line profiles for H$\beta$ and $\mathrm{[O\,III]}\lambda5007$. We then identified an object as a likely Type\,1 AGN if the FWHM of H$\beta$ was greater than that of [O\,III] in 99\;per\;cent (or more) of the realisations and the FWHM of H$\beta$ was greater than 1000\;km\;s$^{-1}$ -- these objects are flagged with \textsc{flag\_type1 = True} in our catalogue. Using this flag, we identify 2,664 likely Type\,1 AGN in our sample. In Figure\;\ref{fig:type1_vel_dist}, we show the H$\beta$ and $\mathrm{[O\,III]}\lambda5007$ FWHM distributions of the likely Type\,1 objects (solid blue contours) that we identified with this flag, alongside the non Type 1 objects (dash-dotted red contours) that we use in the classification scheme. 

\subsection{Evaluating goodness-of-fit}\label{sec:fitting:flag_spec}

To assess the reliability of the spectral fits, we defined a quality-control flag that evaluates the performance of the fitting routine in the regions around the emission lines\footnote{We note that while \texttt{\textsc{WL-SLAYER}} fits the continuum globally and subtracts it across the full spectrum, the quality of the fit in the emission-line regions is what is critical for this analysis.}. Following the approach of \citetalias{arnaudova2025} (also used in e.g. \citealt{smith2012}), we modelled the distribution of the resulting chi-squared ($\chi^2$) values (evaluated at $\pm 4000\ \mathrm{km\ s^{-1}}$ around the Balmer lines and $\pm 2000\ \mathrm{km\ s^{-1}}$ for the forbidden lines) with a $\chi2$ distribution separately for the low-$z$ and high-$z$ samples (independent of the number of Gaussian components used). We then identified the 99th percentile of the best-fitted $\chi2$ distributions, above which spectra are classified as having 'bad' fits and are excluded. Applying these thresholds results in a retained sample (hereafter the classification sample) of 251,413 sources (this includes $\sim90$ per cent of sources for the low-$z$ and 93 per cent for the high-$z$ sample). In our catalogue, these sources can be chosen with \textsc{flag\_spec = True}. All processing steps leading to the final classification sample are summarised in Table~\ref{tab:sample_selection}.

\begin{table}
\centering
\small
\setlength{\tabcolsep}{2pt}
\caption{A summary of the processing steps applied to the cross-matched sample between the catalogue from \citet{hardcastle2023} and DESI DR1.}
\begin{tabular}{lcc}
\HL
\textbf{Processing step} & \textbf{Sources removed} & \textbf{Sources remaining} \\
\HL
LoTSS DR2 matches with DESI & --- &  398,399\\
DESI quality flags& 35,073& 363,326\\
Consistent $z$s& 1,285&362,041\\
Aperture corrections & 54,309& 307,732\\
Sources with $z\leq0.947$ & 28,032& 279,700\\
Excluding Type 1 AGN& 2,664& 277,036\\
Good spectral fits & 25,623& 251,413\\

\HL
\end{tabular}
\label{tab:sample_selection}
\end{table}

\section{Classifications}\label{sec:classifications}

To classify sources as SFGs, RQ AGN, LERGs, and HERGs, we adopt a probabilistic spectroscopic classification method similar to \citetalias{arnaudova2025}. This method uses a combination of a radio-excess diagnostic, the BPT-NII (hereafter BPT), a modified Mass-Excitation ($\mathcal{M}$Ex) diagram and the [\textsc{Oiii}]$\lambda5007$ EW, which we discuss in more detail in the following sections, alongside Monte Carlo realisations to estimate the probability for a source to belong in each of the four categories. 

\subsection{Identifying radio-excess galaxies}\label{sec:radio_excess}

Following \citetalias{arnaudova2025}, radio-excess galaxies (RXGs) are identified in this work using the $L_{\mathrm{H}\alpha}^{\mathrm{corr}}$–$L_{144\mathrm{MHz}}$ diagnostic, where $L_{\mathrm{H}\alpha}^{\mathrm{corr}}$ is the dust-corrected H$\alpha$ luminosity. This relation is a proxy for the linear correlation between star formation rate (SFR) and radio luminosity found for star-forming galaxies (albeit with indication of some stellar mass dependence as found in e.g. \citealt{gurkan2018}; \citealt{smith2021lofar}; \citealt{delvecchio2021}; \citealt{das2024lofar}; \citealt{shennoy2026}) such that sources offset from this relation towards higher radio luminosities are considered to have radio emission due to AGN-related processes such as jets. However, as discussed in \citetalias{arnaudova2025}, the precise location of this demarcation line depends on factors such as aperture corrections. Since our approach to aperture corrections differs from that of \citetalias{arnaudova2025}, we need to re-define the threshold: we adopt the approach of \citetalias{drake2024lofar} (also used in \citetalias{arnaudova2025}), which estimates the peak of the $\log_{10} L_{144\,\mathrm{MHz}}$–$\log_{10} L_{\mathrm{H}\alpha}^{\mathrm{corr}}$ distribution for all sources with a 5$\sigma$ detection in H$\beta$, H$\alpha$ and radio flux using a kernel density estimation (KDE). The left-hand side of this KDE peak (where it is assumed that the SFGs dominate) is then modelled with a Gaussian distribution, and the 99th percentile of the symmetrised Gaussian is taken as the demarcation threshold. Therefore, RXGs identified in this work are sources for which:
\begin{equation}\label{eqn:rexcess}
\log_{10}( L_{144\mathrm{MHz}}/\mathrm{W Hz}^{-1}) > \log_{10}(L_{\mathrm{H}{\alpha}}^{\mathrm{corr}}/ L_{\odot})+ 15.00 (\pm 0.02)
\end{equation}

\noindent where 15.00 corresponds to the 99th percentile of the SFG Gaussian distribution. We note that re-deriving the RX threshold using only sources classified as SFGs according to the BPT diagram (see section~\ref{sec:BPT_MEx}) yields a value of $14.92\pm0.01$. However, as non-radio excess sources dominate our sample (which is the case irrespective of the demarcation line used), we choose to retain equation~\ref{eqn:rexcess} as it provides a more conservative criterion for identifying radio-excess sources and is consistent with previous works (e.g. \citetalias{drake2024lofar}; \citetalias{arnaudova2025}).

This approach is straightforward for the low-$z$ sample where H$\alpha$ lies within the DESI spectral coverage allowing us to directly measure $L_{\mathrm{H}\alpha}$ and calculate the extinction of the H$\alpha$ line ($A_{\rm H\alpha}$) using the Balmer decrement (see \citetalias{arnaudova2025} for details). Beyond $z > 0.483$, applying this method becomes more challenging. Although H$\beta$ remains within the DESI spectral coverage, the lack of consistent high-quality multi-wavelength photometry across the full LoTSS DR2-wide area limits our ability to derive robust extinction values from SED fitting, as was done by \citetalias{arnaudova2025}. However, as part of DESI DR1, several value-added catalogues (VACs) are available, including the AGN Host Galaxies Physical Properties VAC \citep{Siudek2024}, which provides rest-frame $U - V$ and $V - J$ colours for all galaxies derived with the \texttt{\textsc{CIGALE}} SED fitting code. This enables us to use the $UVJ$ diagram (\citealt{Wuyts2007}; \citealt{willaims2009}; \citealt{muzzin2013}), commonly used to distinguish between dusty star-forming and old passively-evolved galaxies, as a way to infer the extinction for the high-$z$ sample. 

\begin{figure}
    \centering
    \includegraphics[width=\linewidth]{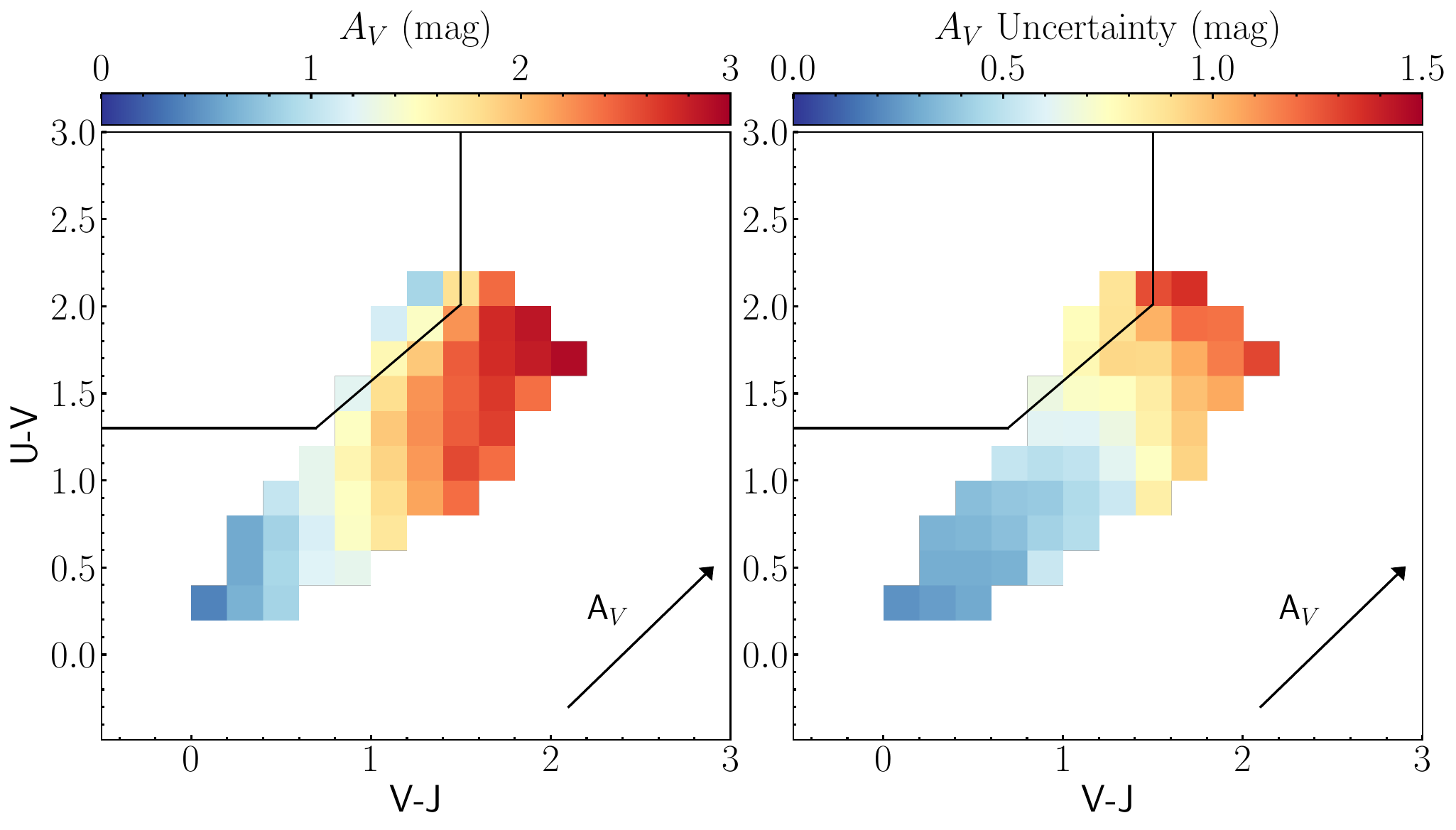}
    \caption{The rest-frame $U-V$ versus $V-J$ (the $UVJ$ diagram) for all sources in our sample that satisfy the \texttt{\textsc{CIGALE}} SED-fitting goodness-of-fit criteria and have a 5$\sigma$ detection in H$\beta$ and H$\alpha$, colour-coded based on the 50th percentile of the Balmer-decrement-derived A$_{V}$ distribution for that bin (left), as well as the spread in those values as estimated by half the difference between the 16th and 84th percentile (right). The solid lines show the adopted divisions between star-forming and quiescent galaxies from \citet{muzzin2013}.}
    \label{fig:UVJ_Av}
\end{figure}

We achieve this by binning the Balmer-derived $V$-band extinction ($A_V$) values\footnote{This is done by calculating the colour excess as $\mbox{E(B-V) = 1.97 log}_{10} \left[ \frac{(\mathrm{H}\alpha/\mathrm{H}\beta)_{\rm obs}}{(\mathrm{H}\alpha/\mathrm{H}\beta)_{\rm int}} \right]$, where $(\mathrm{H}\alpha/\mathrm{H}\beta)_{\rm obs}$ is the observed H$\alpha$-to-H$\beta$ flux ratio and $(\mathrm{H}\alpha/\mathrm{H}\beta)_{\rm int} = 2.86$ is the intrinsic ratio, corresponding to Case~B recombination at $T = 10^{4}\,\mathrm{K}$ and $n_e = 10^{2}\,\mathrm{cm^{-3}}$. We then convert this to $A_V$ using the attenuation curve from \cite{calzetti2000dust}, $A_V = 4.05\,E(B-V)$, and because we ultimately convert $A_V$ back to $A_{\mathrm{H}\alpha}$ using the same attenuation curve, the assumed differential extinction only affects the absolute scaling of $A_V$ in Figure~\ref{fig:UVJ_Av}.}
of the low-$z$ sample (with $5\sigma$ detections in H$\beta$ and H$\alpha$) in the rest-frame $U-V$ and $V-J$ colour space using 0.2\,dex bins. For each bin containing at least 50 objects, we compute the 16th, 50th, and 84th percentiles of the $A_V$ distribution; these are shown in Figure 4. We then assign to each source in the high-$z$ sample an $A_V$ corresponding to the median (50th percentile) of the bin, with an uncertainty estimated as half the difference between the 84th and 16th percentiles. If a source falls outside these bins (affecting only $\sim 10$ per cent of sources), we adopt the values from the nearest bin. This approach works well, as several works (e.g. \citealt{martis2016}; \citealt{fang2018}) have found that the $A_{V}$ increases diagonally, except for sources located in the quiescent region (as denoted by the region marked by the solid black line in Figure \ref{fig:UVJ_Av}) which are associated with low-$A_{V}$ values, and that this trend does not evolve significantly over the redshift range considered in this work. To further validate this approach, we compare the $A_V$ values assigned to the high-$z$ sample with Balmer-derived $A_V$ estimates obtained from H$\gamma$/H$\beta$ measurements in stacked high-$z$ spectra within each $UVJ$ bin (see section \ref{sec:stacks} for details of the stacking procedure). The stacked spectra are fitted using \texttt{\textsc{WL-SLAYER}} in a similar setup as discussed in section~\ref{sec:fitting}.
We qualitatively recover similar trends; however, the combination of the relatively low signal-to-noise of the Balmer lines and the increased difficulty of accurately modelling the stellar
continuum in the stacked spectra limits a more detailed quantitative assessment, which we defer to future work. Nonetheless, this comparison provides empirical support for applying the low-$z$ Balmer-derived $A_V$ assignment method to capture population-level dust trends across $UVJ$ space.
This technique then allows us to infer dust-corrected H$\alpha$ luminosities for the high-$z$ sample as in \citetalias{arnaudova2025} and, in turn, apply equation~\ref{eqn:rexcess} to identify sources exhibiting a radio excess. As discussed in section \ref{sec:build_class}, this technique also provides a route to estimating the extinction for lower-$z$ sources when the Balmer lines are too weak to allow for the Balmer decrement approach to be used.

We note that our radio-excess diagnostic shows a moderate level of disagreement with the absolute $W3$ magnitude-$L_{144\mathrm{MHz}}$ diagnostic of \citet{hardcastle2025}. Specifically, $\sim$10 per cent of sources classified as radio excess using our $L_{\mathrm{H}\alpha}^{\mathrm{corr}}$–$L_{144\mathrm{MHz}}$ diagnostic lie below the radio-excess region defined by the $W3$–$L_{144\mathrm{MHz}}$ diagnostic of \citet{hardcastle2025}, which was developed using the probabilistic spectroscopic classifications of \citetalias{drake2024lofar}. Whilst of course some level of disagreement is expected between different radio excess diagnostics (as indeed there is not perfect agreement between samples of Seyfert galaxies obtained using the BPT and modified $\mathcal{M}$Ex methods, for example), further investigation reveals that the `discrepant' sources are typically bright in $W3$ (observed $W3 < 17$ mag) and exhibit red observed $W2-W3$ (Vega) colours ($W2-W3 > 2$), which are largely absent from the sample used by \citetalias{drake2024lofar}. Taken together, we conclude that differences between the DESI and SDSS spectroscopic selection criteria, as well as the complexity of the mid-infrared SED (which is not fully captured by the power-law extrapolation used to derive the absolute $W3$ magnitude; see \citealt{hardcastle2025} for details) are playing a role here. We defer a fuller investigation to future work, combining the strengths of both photometric and spectroscopic classifications (e.g. Das et al. \textit{in prep.}).

\begin{figure*}
    \centering
    \includegraphics[width=1\textwidth]{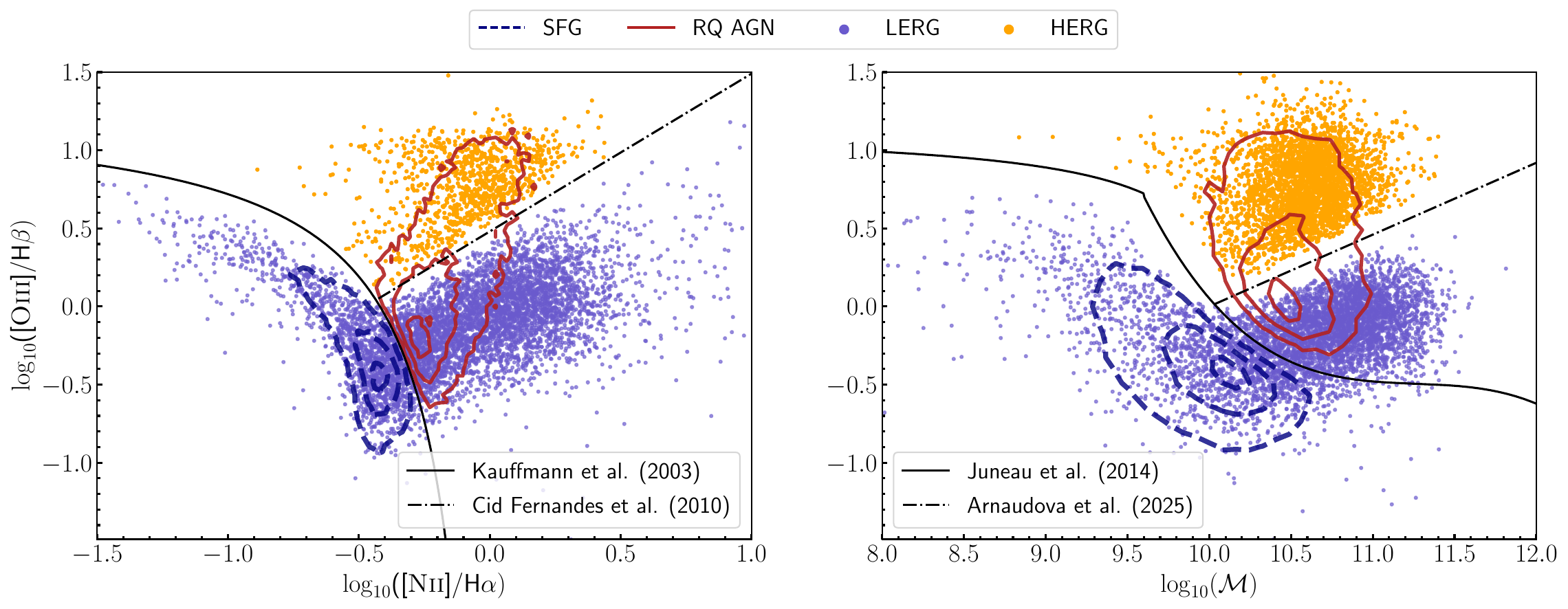}
    \caption{The left panel shows the BPT diagram for sources classified at >90 per cent confidence, where the demarcation line from \citet{kauffmann2003host} separates sources into SFGs (dashed blue contours) and RQ AGN (solid red contours), while the combination of the \citet{kauffmann2003host} line and the diagnostic line from \citet{cidfernandes2010alternative} distinguishes LERGs (purple dots) from HERGs (orange dots). The contours encompass 10, 50 and 90 per cent of the relevant population. Similarly, the right panel presents the $\mathcal{M}$Ex diagram, where the separation between SFGs and RQ AGN is done via the demarcation line from \citet{Juneau2014}, and the division between LERGs and HERGs is based on the \citet{arnaudova2025} demarcation line. We note that the larger gap between the contours in the $\mathcal{M}$Ex diagram compared to the BPT diagram reflects the greater uncertainties in stellar mass measurements relative to those of the [\textsc{Nii}] and H$\alpha$ fluxes.}
    \label{fig:BPT_MEx}
\end{figure*}

\subsection{Identifying radiatively-efficient AGN}\label{sec:BPT_MEx}

To identify radiatively-efficient AGN, we first use the well-known BPT diagnostic diagram (\citealt{baldwin1981}), which uses the emission-line ratios [\textsc{Oiii}]$\lambda5007$/H$\beta$ and [\textsc{Nii}]$\lambda6583$/H$\alpha$ to distinguish galaxies according to their dominant ionisation mechanism. Following \citetalias{arnaudova2025}, this diagram identifies radiatively-efficient AGN among sources without a radio excess (see section \ref{sec:radio_excess}), namely classifying them as either SFGs or RQ AGN\footnote{We note that this scheme classifies sources (without radio excess) that fall within the BPT LINER region as RQ AGN, although previous works have shown that emission in this region may not always be AGN-driven (e.g. \citealt{statsinka2008}). We adopt this classification approach to maintain consistency with previous spectroscopic radio-source classifications (e.g. \citetalias{drake2024lofar}; \citetalias{arnaudova2025}), but provide the full set of classification probabilities, including the probabilistic locations within the diagnostic diagrams, allowing users to adopt alternative treatments of these sources.}, using the empirical demarcation line from \citet{kauffmann2003host}. Sources exhibiting a radio excess are instead separated into LERGs and HERGs using, in addition to the \cite{kauffmann2003host} line, the division from \citet{cidfernandes2010alternative}, which traces the transition between low- and high-ionisation systems in a manner consistent with the Excitation Index (EI; \citealt{Buttiglione2010optical}; \citealt{best2012on}), a parameter commonly used to distinguish between these two classes. The location of all four classes -- SFGs, RQ AGN, LERGs, and HERGs -- in this BPT classification scheme at >90 per cent confidence (see details in section \ref{sec:build_class}) are shown in the left panel of Figure~\ref{fig:BPT_MEx}.

At higher redshifts ($z>0.483$), where H$\alpha$ and [\textsc{Nii}]$\lambda6583$ are no longer observable in the optical range, we turn to the $\mathcal{M}$Ex diagram. This diagnostic builds upon the original Mass-Excitation diagram introduced by \citet{Juneau2011, Juneau2014}, in which the $x$-axis of the BPT diagram is replaced by stellar mass. The modified version incorporates a redshift-dependent offset which is necessary to account for the evolution of the mass–metallicity relation, as discussed in recent studies (e.g. \citealt{henry2021}; \citealt{cleri2023}; \citetalias{arnaudova2025}). This correction shifts the dividing lines presented by \cite{Juneau2014}, which allow us to separate SFGs and RQ AGN, to higher stellar masses at increasing redshift (see \citetalias{arnaudova2025} for details), and is defined as:
\begin{equation}
    \log_{10}\mathcal{M} = \log_{10}M - [1-\mathrm{exp}(-1.2z)].
\end{equation}

\noindent Here, $M$ refers to the stellar mass, taken from DESI's AGN Host Galaxies Physical Properties VAC, while $\mathcal{M}$ denotes the modified stellar mass.
To further separate the RXGs into LERGs and HERGs, we use, in addition to the \cite{Juneau2014} line, the EI-based demarcation line from \citetalias{arnaudova2025}, which allows us to classify sources with the $\mathcal{M}$Ex diagram into SFGs, RQ AGN, LERGs, and HERGs (hereafter the $\mathcal{M}$Ex scheme). The resulting $\mathcal{M}$Ex classification scheme for sources above the 90 per cent threshold is shown in the right panel of Figure \ref{fig:BPT_MEx}. Note that in both the BPT and $\mathcal{M}$Ex diagrams, LERGs overlap with the 'LINER' AGN region (above the \citealt{kauffmann2003host}/\citealt{Juneau2014} line but below that of \citealt{cidfernandes2010alternative}/\citetalias{arnaudova2025}) and also with the SFG region. As weak emission-line AGN, LERGs can have their emission-line properties dominated by star formation in the host galaxy, but the radio excess reveals the presence of AGN activity.

\subsection{Recovering LERGs with low-significance lines}

\begin{figure}
    \centering
    \includegraphics[width=1\columnwidth]{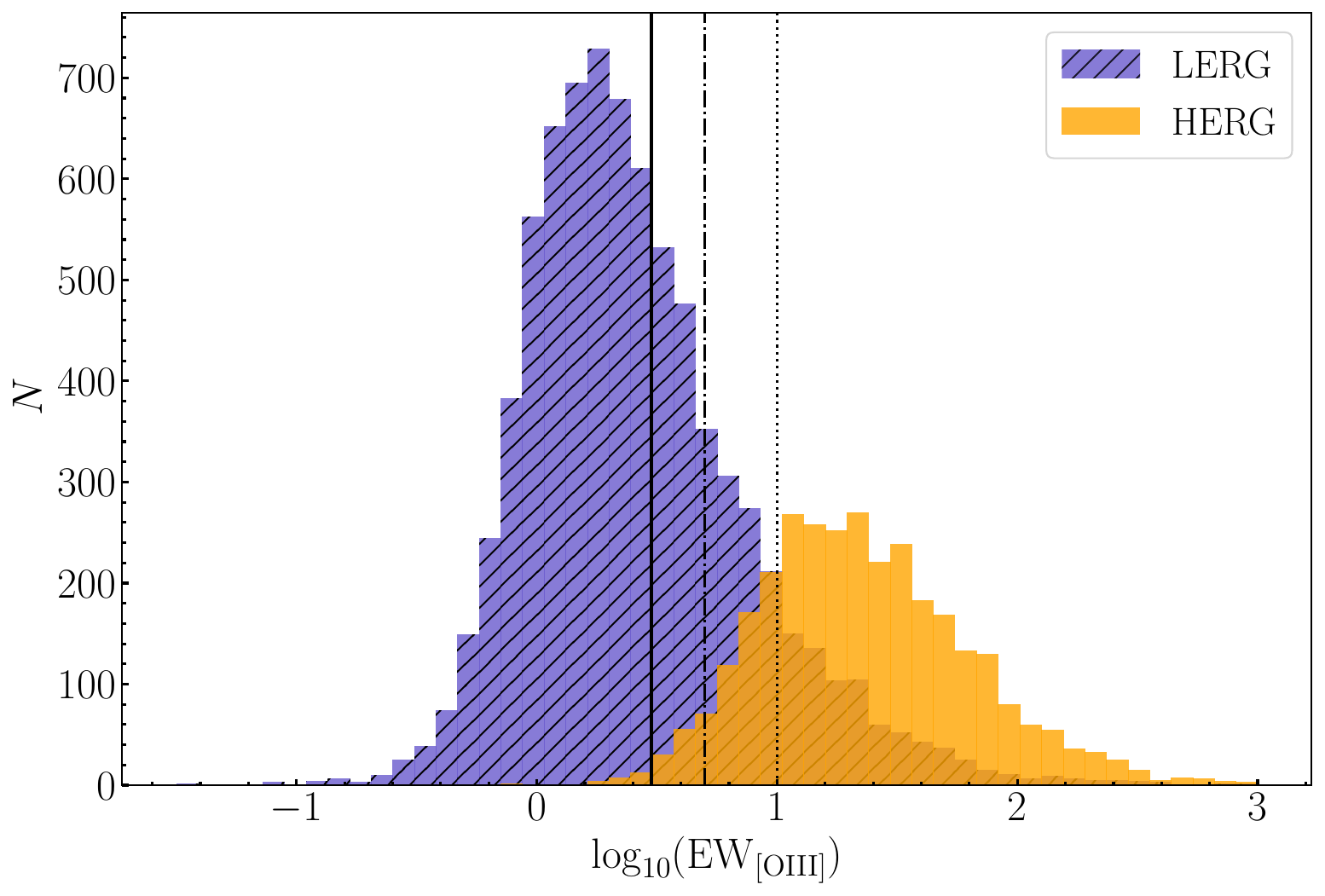}
    \caption{ The distribution of EW$_{\rm[OIII]}$ for sources classified at >90 per cent confidence as LERGs and HERGs using the BPT and $\mathcal{M}$Ex classification schemes. The vertical solid, dashed and dotted lines denote the suggested divisions from \citet{laing1994} at EW$_{\rm [OIII]}$=3\AA, \citet{best2012on} at EW$_{\rm [OIII]}$=5\AA, and \citet{tadhunter1998} at EW$_{\rm [OIII]}$=10\AA, respectively.}
    \label{fig:EW}
\end{figure}

In addition to the BPT and $\mathcal{M}$Ex diagnostics, we aim to improve the classification of LERGs, particularly those associated with low signal-to-noise emission lines that may be missed by these two schemes. To recover such sources, we use the equivalent width of [\textsc{Oiii}]$\lambda5007$ (EW$_{\rm [OIII]}$). Previous studies (e.g. \citealt{laing1994}; \citealt{tadhunter1998}; \citealt{best2012on}) have shown that EW$_{\rm [OIII]}$ provides an effective separation between LERGs and HERGs, and this also holds for the subset of sources already classified using the BPT/$\mathcal{M}$Ex diagnostics (see Figure \ref{fig:EW}). The transition between LERGs and HERGs appears around 10 \AA, with the majority of LERGs having EW$_{\rm [OIII]} < 3$\AA, consistent with the division proposed by \citet{laing1994}. We therefore adopt the more stringent division from \citet{laing1994} to recover additional LERGs within the radio-excess population.

\subsection{Probabilistic classifications}\label{sec:build_class}

To construct the full probabilistic classification scheme, we generate 1,000 Monte Carlo realisations in which we perturb the stellar masses, emission-line fluxes, radio fluxes and the EW$_{\rm [OIII]}$ using their respective uncertainties. For each realisation, we apply the four diagnostics discussed in the previous section to derive the probability of each source belonging to one of the four classes for the entire classification sample, including sources with weak or undetected emission lines.

For the RX diagnostic, we calculate the extinction values for each Monte Carlo realisation. However, as noted in \citetalias{arnaudova2025}, low-significance emission lines can sometimes yield negative fluxes, producing unphysical Balmer-derived extinction values for the low-$z$ sample. In such cases, we use the $UVJ$ diagram in the same manner as for the high-$z$ sample to assign $A_{\rm H\alpha}$ when H$\beta$, H$\alpha$, or both are negative. To track how often this procedure is applied and to allow users to decide whether to include these sources we flag the fraction of realisations for which the $UVJ$- based substitution is used (\textsc{flag\_$A_{V}$}).
Negative fluxes can also propagate into the derived quantities used in the diagnostics, such as the dust-corrected H$\alpha$ luminosity ($L_{\mathrm{H}\alpha}^{\mathrm{corr}}$) and radio luminosity ($L_{\mathrm{144MHz}}$) in the RX diagram, as well as the logarithmic line ratios in the BPT and $\mathcal{M}$Ex diagrams, resulting in undefined or extreme values. To address this, we apply the following conditions: (i) when the numerator is positive but the denominator negative, we assign an unambiguously large value on the relevant axis in the RX, BPT, or $\mathcal{M}$Ex diagram; (ii) when the numerator is negative but the denominator positive, we assign an unambiguously small value; and (iii) when both are negative, the realisation is considered invalid, as no classification can be assigned. We then again record the fraction of affected realisations as three warning flags (\textsc{RX\_warning\_frac}, \textsc{BPT\_warning\_frac}, and \textsc{$\mathcal{M}$Ex\_warning\_frac}), allowing the end user to determine how to treat these cases.

In this work, we construct the final classification using a hierarchical approach. First, we apply the BPT scheme (that is the combination of the RX and BPT diagram) to all sources where H$\alpha$ and [\textsc{Nii}]$\lambda6583$ are available, adopting a high-reliability threshold of 90 per cent for each class. Sources for which this BPT classification falls below this threshold, or which lack H$\alpha$ and [\textsc{Nii}] measurements (i.e. in the high-$z$ sample), are then classified using the $\mathcal{M}$Ex scheme (that is the RX and $\mathcal{M}$Ex diagram), also with a 90 per cent reliability threshold. Following \citetalias{arnaudova2025}, we additionally require \textsc{RX/BPT/$\mathcal{M}$Ex\_warning\_frac} < 0.10, limiting cases where realizations produce negative fluxes to at most 10 per cent, and retain all low-$z$ sources regardless of the \textsc{flag\_$A_{v}$}. Sources that remain unclassified but exhibit a radio excess are further examined using the EW$_{\rm [OIII]}$ criterion: those with more than 90 per cent of Monte Carlo realisations having EW$_{\rm [OIII]} < 3$\AA~ (\textsc{EW\_frac} > 0.90) are further classified as LERGs. Here, we also allow cases where the H$\alpha$ flux can be negative up to 50\% (that is, \textsc{RX\_warning\_frac<0.5}) as this behaviour is expected for weak-lined LERGs whose H$\alpha$ emission may be consistent with zero within the uncertainties while still remaining securely in the radio-excess regime. We note that this procedure increases the number of LERGs by a factor of $\sim 5$. Considering also the quality flags discussed in section~\ref{sec:fitting:type1_flag}~and~\ref{sec:fitting:flag_spec} (i.e. sources with \textsc{flag\_spec} set, and not flagged as Type 1 AGN by \textsc{flag\_type1}), we find 68,820 SFGs, 32,288 RQ AGN, 35,210 LERGs (7,765 classified via the BPT/$\mathcal{M}$Ex schemes, 15,520 via EW$_{\rm [OIII]}$ with \textsc{RX\_warning\_frac}<0.1, and 11,771 via EW$_{\rm [OIII]}$ with $0.1<\textsc{RX\_warning\_frac}<0.5$), 3,085 HERGs and 112,010 unclassified sources. 

We emphasise that the adopted 90 per cent reliability threshold is intentionally conservative and is designed to provide fiducial classification. Users of this catalogue may construct alternative samples tailored to specific science goals, for example by lowering the reliability threshold (e.g. to 70 per cent) or by assigning each source to the class with the maximum probability, rather than applying a hard cut. Such approaches increase completeness at the expense of reliability and can significantly reduce the fraction of unclassified sources. The large number of unclassified sources in this work is therefore an expected consequence of this choice. We can, however, further characterise these unclassified sources by considering whether a radio excess is present. We identify approximately 7,959 sources that exhibit a radio excess at >90 per cent (that is $P(\mathrm{RX})$>0.90 and \textsc{RX\_warning\_frac} < 0.50) but cannot be robustly separated into LERGs or HERGs (uncertain radio-excess objects or RX\_Unc), even after applying the EW$_{\rm [OIII]}$ criterion. Similarly, there are 24,737 sources that securely do not have a radio excess ($P(\mathrm{RX})$<0.10 and \textsc{RX\_warning\_frac} < 0.10\footnote{We apply the relaxed \textsc{RX\_warning\_frac} threshold only when selecting RX candidates. For these sources, realisations with negative H$\alpha$ fluxes (i.e. low-S/N H$\alpha$ emission) are consistent with the expected properties of weak-lined LERGs. In contrast, for non-RX sources we adopt the stricter requirement of \textsc{RX\_warning\_frac}<0.10 to ensure that the absence of a radio excess is not driven by poorly constrained H$\alpha$ measurements.} for which the available diagnostics do not allow a confident separation between SFGs and RQ AGN (uncertain non-radio-excess objects or $\overline{\rm{RX}}$\_Unc) at this threshold (see also Table \ref{tab:target_classification} for a detailed breakdown of the classes across the DESI surveys). The remaining 79,314 unclassified sources (or Unc.) are generally located near the boundaries of the classification criteria and/or are at higher redshifts where the spectral S/N is lower (see section \ref{sec:results}), such that they cannot be robustly classified using the available spectroscopic and radio diagnostics at the adopted 90 per cent reliability threshold. Relaxing the reliability requirement from 90 to 70 per cent would reduce the number of unclassified sources from 79,314 to 43,289, while increasing the numbers of classified SFGs, RQ AGN, LERGs, and HERGs to 77,010, 46,498, 49,412 and 5,501, respectively. Future work (Das et al. \textit{in prep.}) will investigate the inclusion of additional information from broad-band photometry and joint photometric–spectroscopic SED fitting within a Bayesian framework, moving beyond the frequentist Monte Carlo approach adopted here. Such methods offer a natural way to incorporate informative priors that may help to further reduce the unclassified fraction of sources. 

We note that the hierarchical classification adopted in this work results in 1,242 (less than 1 per cent of the classification sample) sources changing class between the 70 and 90 per cent reliability thresholds. This occurs when a source has a BPT classification probability above 70 per cent but below 90 per cent, in which case the classification is instead determined by the $\mathcal{M}$Ex or EW$_{\rm [OIII]}$ diagnostic at the higher reliability threshold. In particular, 72 sources classified as LERGs at 70 per cent threshold are instead classified as HERGs at 90 per cent, and 28 vice versa. Similarly, 1,010 RQ AGN at the 70 per cent threshold are instead classified as SFGs at 90 per cent, while 132 sources change from SFGs to RQ AGN. Given the small number of affected sources, this does not significantly impact the results presented in the following sections. Nevertheless, we note that these sources can be identified in the released catalogue and removed if a fully self-consistent hierarchical classification is preferred.

\begin{table}
\centering
\small
\setlength{\tabcolsep}{1.5pt}
\caption{A breakdown of source classifications at $>90$ per cent confidence across the DESI surveys described in Section \ref{sec:desi data}, including both radio-excess sources that cannot be classified as LERGs or HERGs (RX\_Unc) and non-radio-excess sources that cannot be classified as SFGs or RQ AGN ($\overline{\rm{RX}}$\_Unc).}
\begin{tabular}{lcccccccc}
\HL
\textbf{Survey} & \textbf{SFG} & \textbf{RQ AGN} & \textbf{LERG} & \textbf{HERG} & \textbf{RX\_Unc} &\textbf{$\overline{\rm{RX}}$\_Unc} & \textbf{Unc} &\textbf{Total}\\
\HL
All & 68,820 & 32,288 & 35,210 & 3,085 & 7,959 & 24,737 & 79,314 & 251,413\\
BGS & 63,893 & 28,533 & 21,663 & 1,246 & 1,986 & 21,782 & 41,881 & 180,984\\
LRG & 1,075 & 5,445 & 18,496 & 1,843 & 5,738 & 3,528 & 41,341 & 77,466\\
ELG & 30 & 19 & 105 & 29 & 177 & 161 & 402 & 923\\
QSO & 45 & 46 & 140 & 36 & 176 & 186 & 341 & 970\\
MWS & 41 & 93 & 94 & 4 & 7 & 31 & 82 & 352\\
SCND & 2,312 & 1,395 & 639 & 56 & 193 & 615 & 1,494 & 6,704\\

\HL
\end{tabular}
\label{tab:target_classification}
\end{table}

\section{Results}\label{sec:results}
In this section, we compare the results of the classification scheme discussed in the previous section with those obtained in \citetalias{drake2024lofar} using SDSS spectroscopy, and examine the selection effects introduced by DESI's targeting strategy.

\subsection{Comparison with SDSS}
\subsubsection{Direct Comparison}

To make a direct comparison with the probabilistic classification scheme from \citetalias{drake2024lofar}, we identify the sources in common with the \citetalias{drake2024lofar} catalogue\footnote{Available from \url{https://lofar-surveys.org/dr2_release.html}} and our parent sample using the LoTSS Object identifier. In doing so, we identify 29,996 common sources between DESI DR1 and SDSS, of which 26,266 are included in our classification sample once the type 1 AGN and `bad' spectral fits are removed (see section \ref{sec:fitting:type1_flag} and \ref{sec:fitting:flag_spec}). To construct the 90 per cent confidence sample from \citetalias{drake2024lofar}, we select sources that have $>90$ per cent probability in each class in the catalogue — SFGs, RQAGN, LINELERG (emission-line LERGs; sources with measured BPT emission lines), and HERG — while also satisfying the recommended quality conditions (\textsc{zscore} < 2.5 and \texttt{$\rm CLASS\_z\_WARNING == 0$}). This results in 5,456 SFGs, 3,070 RQ AGN, 2,141 LERGs and 64 HERGs, according to \citetalias{drake2024lofar}.

\begin{figure}
    \centering
    \includegraphics[width=1\columnwidth]{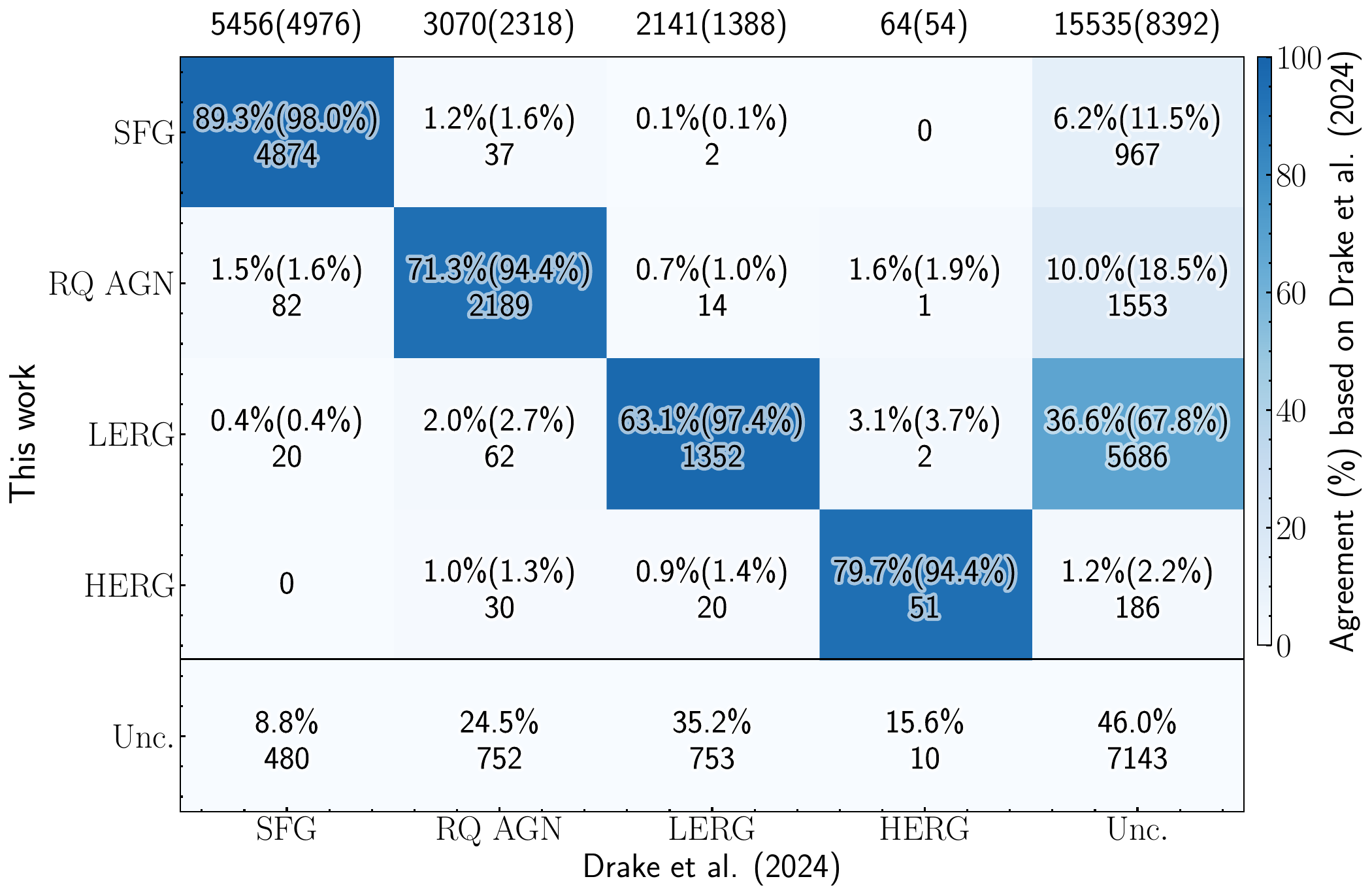}
    \caption{Confusion matrix comparing the spectroscopic classifications with >90 per cent confidence between this work (rows) and \citetalias{drake2024lofar} (columns) for the common DESI and SDSS LOFAR sources. Each cell represents the level of agreement based on \citetalias{drake2024lofar} in percentages, where the number of sources in each cell is also included. The percentages in brackets are calculated for the subset of sources with classifications (that is, not including the unclassified sources) in both works and is represented by the colourbar. The total number of sources in each of the \citetalias{drake2024lofar} classes is shown at the top, with the bracketed numbers indicating the number of \citetalias{drake2024lofar} sources with classifications in this paper.}
    \label{fig:conf_Dr24}
\end{figure}

A detailed comparison between the two schemes is shown in Figure \ref{fig:conf_Dr24} in the form of a confusion matrix, where the results are presented as percentages, showing the
level of agreement when considering \citetalias{drake2024lofar} classifications as the true values. It can be seen that most of the cases where the classifications disagree are because one classification scheme is able to produce a > 90 per cent classification confidence and the other is not, leading to the source being deemed as unclassified. To demonstrate this, we present the subset that meets the 90 per cent threshold for both works in brackets. When restricting the comparison to this subset, the agreement increases to $\sim95$ per cent in each class. There are a number of reasons for discrepancies in whether a >90 per cent confidence classification could be obtained. First of all, we adopt aperture corrections from \texttt{\textsc{FastSpecFit}}, while \citetalias{drake2024lofar} only considers sources with small optical sizes (having a half-light radius of $R_{50}<15$ arcsec) and applies no aperture correction, which results in a different RX demarcation line. Their selection of only small optical sources can also bias the sample toward lower-mass systems, particularly at low redshift. Secondly, we apply a different demarcation line than \citetalias{drake2024lofar} to separate LERGs and HERGs in the BPT diagram, where \citetalias{drake2024lofar} adopt the \cite{kauffmann2003host} line, while we use the one from \cite{cidfernandes2010alternative} which \citetalias{arnaudova2025} found to better follow the EI diagnostic. In addition, we use the warning fractions to account for negative fluxes in individual MCMC realisations, which are propagated into the classification results (see Section~\ref{sec:build_class}). Finally, we use additional diagnostics such as the $\mathcal{M}$Ex diagram and the EW$_{\rm [OIII]}$, which allow us to classify significantly more sources, as shown in the fifth column. This is particularly evident for LERGs, where we recover nearly three times as many sources compared to \citetalias{drake2024lofar} that would otherwise be missed. These, combined with the inherent randomness introduced by noise of the different spectra and the Monte Carlo realisations used to estimate classification probabilities, which may lead to sources being just over the 90 per cent threshold in one dataset and just below the other, explain the inconsistencies between the two works.

\begin{figure*}
    \centering
    \includegraphics[width=0.94\textwidth]{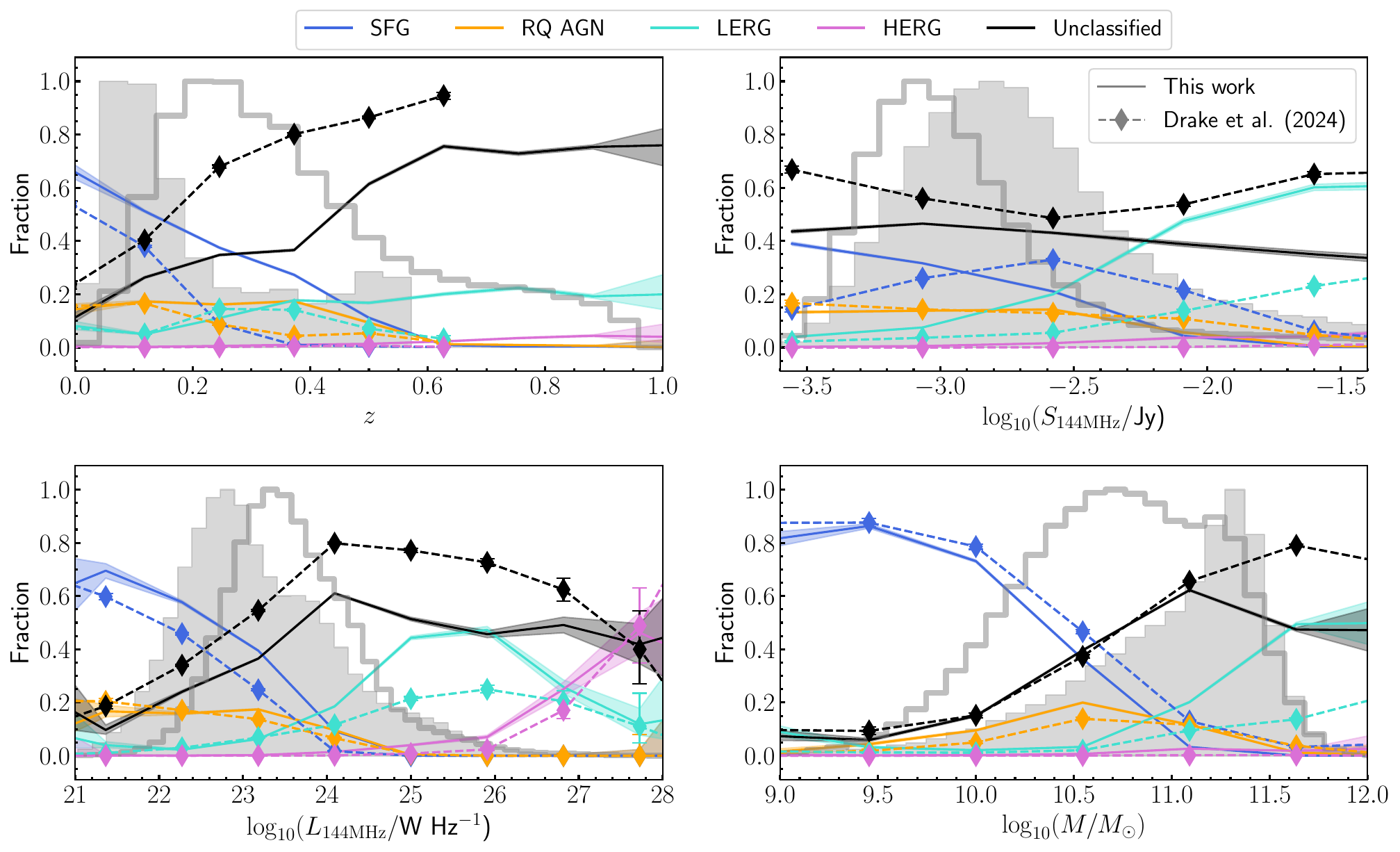}
    \caption{The distributions of sources classified using our $>90$ per cent reliability threshold (solid lines and step histograms) and those from \citetalias{drake2024lofar} (dashed lines and filled histograms) as a function of redshift (top left), 144~MHz flux density (top right), radio luminosity (bottom left), and stellar mass (bottom right). Each panel shows the distributions for the different classes: SFGs (blue), radio-quiet AGN (orange), LERGs (turquoise), HERGs (purple), and unclassified sources (black), as indicated in the legend at the top. The error bars and shaded regions accompanying each panel represent binomial uncertainties. }
    \label{fig:demographics1}
\end{figure*}

\begin{figure*}
    \centering
    \includegraphics[width=0.94\textwidth]{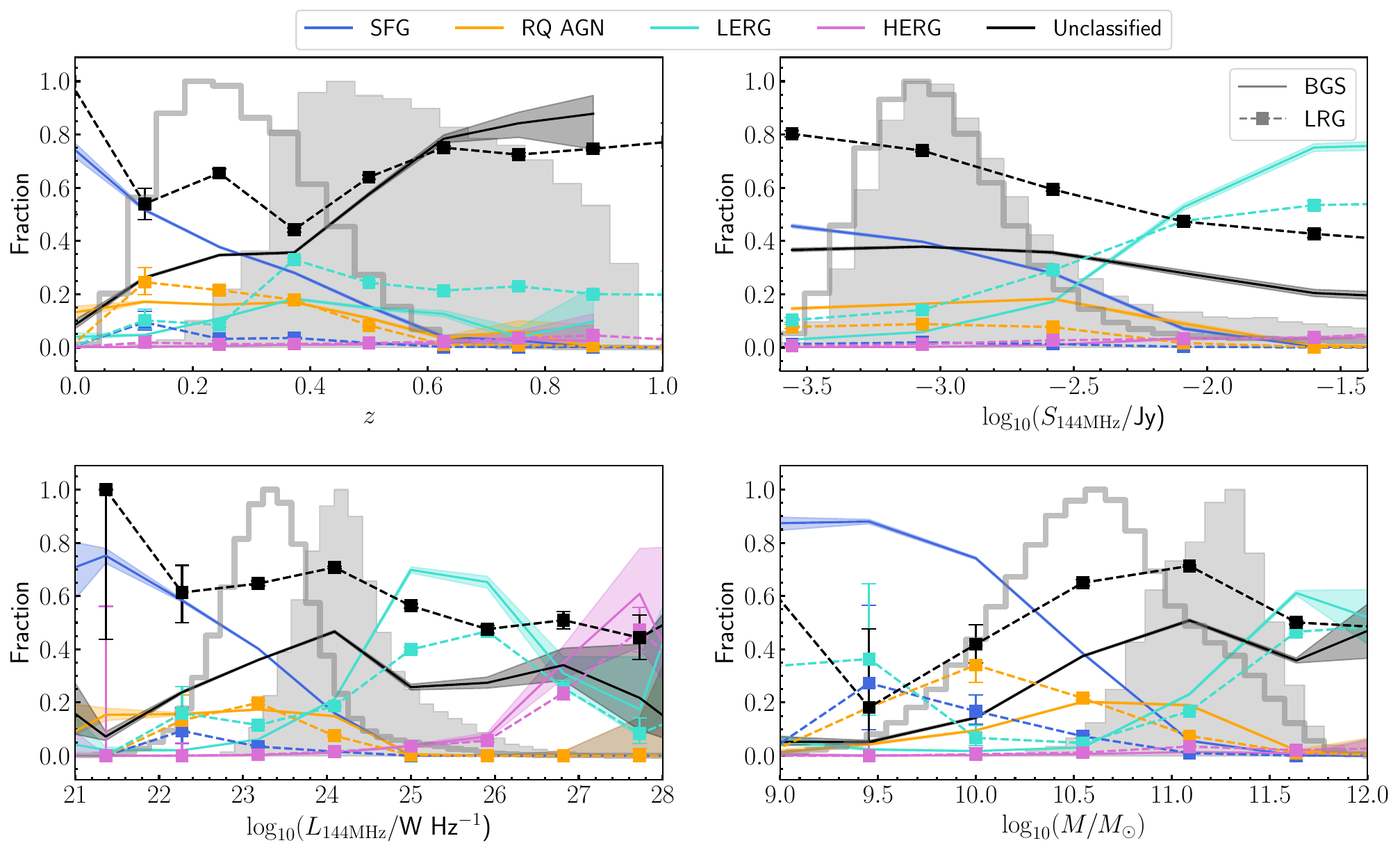}
    \caption{The influence of optical selection criteria on the sampling of the faint radio source population, illustrated by the distributions of sources classified above our $>90$ per cent reliability threshold in the Bright Galaxy Survey (BGS; solid lines and step histograms) and the Luminous Red Galaxy sample (LRG; dashed lines and filled histograms). Panel layout, line styles, and shading are the same as in Figure \ref{fig:demographics1}. }
    \label{fig:demographics2}
\end{figure*}

\subsubsection{Overall Comparison}

To make an overall comparison between the two probabilistic classifications, we examine the demographics of the classes based on the redshift, 144~MHz flux density, radio luminosity and stellar mass, as shown in Figure \ref{fig:demographics1}. In each panel, our classifications are shown by the solid lines, while those from \citetalias{drake2024lofar} are shown by the dashed lines. Here, for improved statistics, we use the entire sample from \citetalias{drake2024lofar} (not just the sources in common with DESI), where applying the 90 per cent threshold and recommended flags as in the previous section gives us 38,704 SFGs, 18,654 RQ AGN, 12,110 LERGs, 359 HERGs and 82,528 unclassified sources, classified according to \citetalias{drake2024lofar}. 

The top-left panel shows the redshift distributions, where we can see that our unclassified fraction is significantly smaller than that of \citetalias{drake2024lofar} across the overlapping redshift range. This improvement probably reflects the combined effect of using multiple diagnostic diagrams, robust spectral fitting techniques, and an improved treatment of the Monte Carlo realisations. At higher $z$, our unclassified fraction increases as our sample becomes both fainter and increasingly dominated by DESI's Luminous Red Galaxies (LRGs; see section~\ref{sec:desi_selection}). These systems are typically passive and often lack the strong emission lines necessary for confident classification. We also observe a higher fraction of SFGs and RQ AGN, particularly in the $0.2 < z < 0.4$ range. This partly results from the same improvements that allow more sources to be classified, but may also reflect selection effects: in this redshift regime, our sample includes many sources from the DESI Bright Galaxy Survey (BGS), which targets fainter optical and thus less massive galaxies compared to SDSS (as demonstrated by the stellar mass distribution of BGS sources in Figure \ref{fig:demographics2}) that, when detected by LOFAR, are more likely to be associated with star formation activity (see further details in section~\ref{sec:desi_selection}).

In the top-right panel, the flux density distribution shows that both our method and that of \citetalias{drake2024lofar} maintain a relatively flat unclassified fraction across the full range of 144~MHz fluxes, indicating that the ability to classify sources is mostly governed by the spectroscopic data quality. We do, however, observe a modest decrease in our unclassified fraction at the highest flux densities ($S_{\rm 144,MHz}\gtrsim2.2$~mJy), accompanied by a corresponding increase in the fraction of sources classified as LERGs in this work, suggesting that our classification scheme may be more effective at recovering radio-excess AGN at higher radio flux densities. This LERG fraction is significantly larger than that in \citetalias{drake2024lofar}, likely due to the inclusion of the EW$_{\rm [OIII]}$ diagnostic. We also identify a larger fraction of SFGs at the lowest flux densities, whereas above $S_{\rm 144,MHz}\simeq1.3$~mJy the SFG fraction is larger in \citetalias{drake2024lofar}. The distributions for RQ AGN and HERGs remain broadly consistent between the two works. These differences likely reflect a combination of classification methodology and survey selection effects (DESI versus SDSS), rather than intrinsic physical differences.

By contrast, the radio luminosity (bottom left) and stellar mass (bottom right) distributions indicate that the underlying physical properties of the classified populations remain broadly consistent between the two classification schemes, despite the differences seen in redshift and flux density. The SFGs dominate at $L_{144\rm MHz}\lesssim10^{24}$W\,Hz$^{-1}$, while LERGs and HERGs become increasingly prevalent at higher radio luminosities. Similarly, SFGs are predominantly associated with lower stellar mass systems, whereas the AGN populations are mostly hosted by more massive galaxies ($M_{*}\gtrsim10^{10}M_{\odot}$).
We do, however, identify a larger fraction of LERGs at $L_{144\rm MHz}\lesssim10^{24-27}$W\,Hz$^{-1}$ and $M_{*}\gtrsim10^{11}M_{\odot}$ compared to \citetalias{drake2024lofar} which again is due to the additional weak-lined LERGs identified through the EW$_{\rm [OIII]}$ diagnostic.

\subsection{DESI Selection}\label{sec:desi_selection}

As discussed in section \ref{sec:desi data}, DESI consists of multiple surveys with distinct selection functions, which influence the populations we classify. As shown in Table \ref{tab:target_classification}, the majority of our classified sources come from the BGS and LRG samples. The BGS targets low-redshift ($0<z<0.6$), relatively bright galaxies, including a mix of star-forming and passive systems, whereas the LRG sample selects high-mass, red, passively evolving galaxies at higher redshift ($0.4<z<1.1$). Therefore, to explore whether these selection strategies impact our results, we examine the demographics of classifications split by the two surveys, again considering redshift, 144~MHz flux density, radio luminosity, and stellar mass, as shown in Figure \ref{fig:demographics2}. In each panel, BGS classifications are represented by solid lines, and LRG classifications by dashed lines.

The top-left panel shows the redshift distributions of the BGS and LRG subsets. The LRG sample exhibits a persistently high fraction of unclassified sources ($>50\%$) across all redshifts, reflecting the passive nature of these galaxies and the lack of strong emission lines required for classification. In contrast, the BGS sample is largely classified at low redshift, with the unclassified fraction increasing only beyond $z\gtrsim0.5$, where the number of BGS targets declines sharply. As expected, BGS sources are dominated by SFGs at all redshifts, whereas LRGs contain very few SFGs. RQ AGN and LERGs are similarly represented in both samples.

The top-right panel, showing the 144~MHz flux density distribution, reveals a more pronounced decrease in the unclassified fraction towards higher flux densities, especially for the LRG sample, accompanied by a corresponding increase in the fraction of sources classified as LERGs. This trend further supports the idea that our classification scheme is more effective at recovering radio-excess AGN at higher radio flux densities, particularly within the LRG population.
The BGS sample seems to be dominated by SFGs and RQ AGN at low fluxes, with LERGs becoming more common above $S_{\rm144MHz} \gtrsim -3.2$~mJy. The LRG sample, on the other hand, is dominated by LERGs across the entire radio flux range, with some contribution of RQ AGN at lower radio fluxes and HERGs at $S_{\rm144MHz} \gtrsim -2.2$~mJy. The radio luminosity distributions in the bottom-left panel highlight the expected trends observed in the previous section: radio-excess sources dominate above $L_{144\rm MHz}=10^{24}$~W~Hz$^{-1}$, while SFGs and RQ AGN lie primarily below this threshold. Notably, the LERG distribution in the LRG sample does not decrease strongly towards low radio powers, revealing a population of weaker radio AGN. These objects also appear to be associated with lower stellar masses, as can be seen in the bottom-right panel. A similar behaviour is observed for the RQ AGN within the LRG sample, which tend to reside in lower-mass galaxies than those in the BGS population.

Overall, these trends illustrate that DESI’s targeting strategy influences the observed class distributions. These selection effects must therefore be taken into account when interpreting the physical properties and relative prevalence of radio source populations in the DESI–LOFAR sample.

\section{The properties of LERGs and HERGs}\label{sec:properties_lergs_hergs}

\begin{figure}
    \centering
    \includegraphics[width=1\columnwidth]{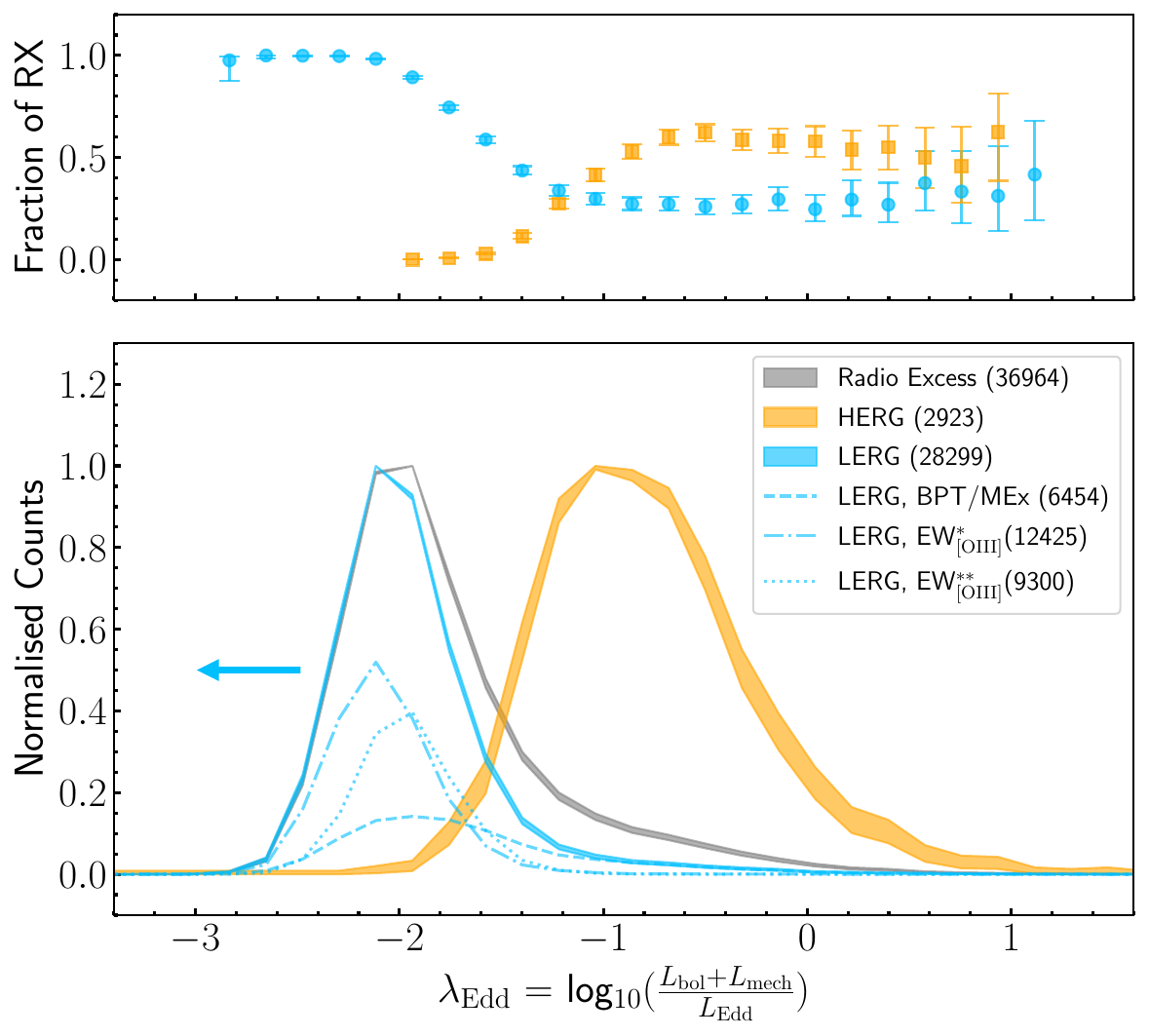}
    \caption{The log Eddington-scaled accretion rate distributions for LERGs (blue), HERGs (orange), and radio-excess sources (grey; including objects that cannot be confidently assigned to either class) that satisfy our $>90$ per cent threshold. For the LERG population, the distributions are additionally separated by classification method: the dashed line indicates sources classified via the BPT/$\mathcal{M}$Ex diagnostics, while the dot-dashed (*) and dotted (**) lines indicate sources classified using the EW$_{\rm [OIII]}$ method with \textsc{RX\_warning\_frac} < 0.10 and 0.10 < \textsc{RX\_warning\_frac} < 0.50, respectively. The arrow indicates that the LERG accretion rates should be considered upper limits due to the $3\sigma$ upper limit on the [\textsc{Oiii}] emission, while the shaded regions indicate binomial uncertainties. The number of sources in each class is given in the legend at the upper right; these numbers differ from Table \ref{tab:target_classification} as not all sources have reliable stellar mass estimates. The upper panel shows the fraction of radio-excess sources classified as LERGs (blue) and HERGs (orange), with error bars again representing binomial uncertainties. Only bins containing at least five sources are included.}
    \label{fig:lambda_edd}
\end{figure}

\begin{figure*}
    \centering
    \includegraphics[width=1\textwidth]{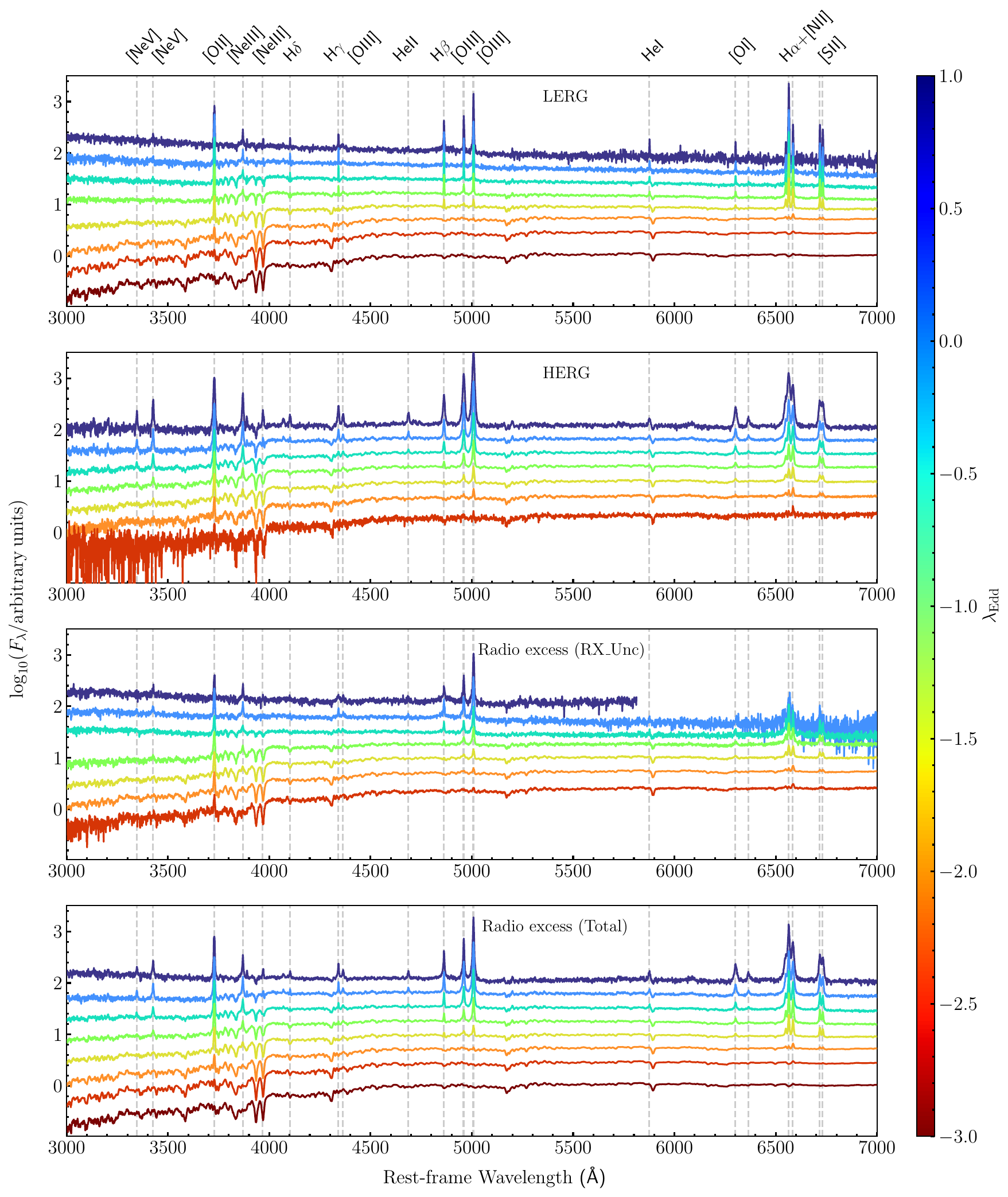}
    \caption{The stacked spectra of LERGs (top panel), HERGs (second panel), radio-excess sources that cannot be classified at $>90$ per cent confidence into the two classes (or RX\_Unc, third panel) and the whole radio excess population (bottom panel) shown as a function of log Eddington-scaled accretion rate, as indicated by the colourbar. Each stack represents the median of all spectra within a given accretion-rate bin in the rest frame, as indicated by the colour bar. Prominent emission lines are marked with dashed grey lines and labelled in the top panel. }
    \label{fig:lambda_edd_stacks}
\end{figure*}

\begin{figure}
    \centering
    \includegraphics[width=1\columnwidth]{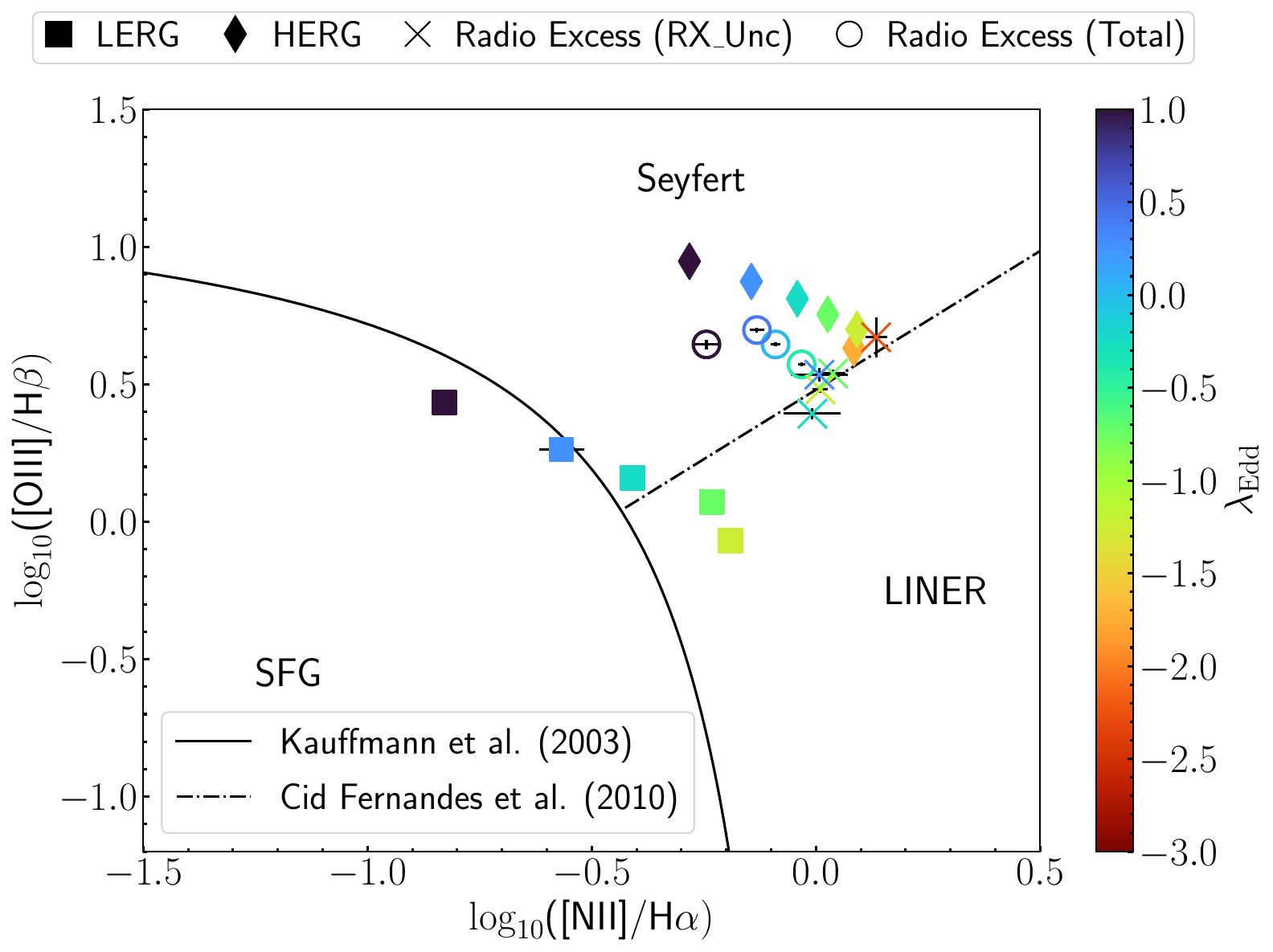}
    \caption{The BPT diagram for the stacked spectra of LERGs (squares), HERGs (diamonds), radio-excess sources that remain unclassified (or RX\_Unc; cross), and the total radio-excess sources (circles), colour coded by the log Eddington-scaled accretion rate. The diagnostic lines delineating the SFG, LINER and Seyfert region are from \citet{kauffmann2003host} and \citet{cidfernandes2010alternative} and are denoted in the legend.}
    \label{fig:BPT_lambda_edd}
\end{figure}

\subsection{Eddington-scaled accretion rates}\label{sec:accretion_rates}

Using a similar classification method, \citetalias{arnaudova2025} showed that the Eddington-scaled accretion rates of LERGs and HERGs remain distinct up to $z=1$, contrary to some recent findings in the literature (e.g. \citealt{whittam2022}). \citetalias{arnaudova2025} attributed this result to the use of photometric methods in those literature studies, showing that spectroscopic classifications break the degeneracy. However, the sample of \citetalias{arnaudova2025} was relatively small and did not allow a detailed investigation of the spectral properties of LERGs and HERGs as a function of accretion rate. This is of particular interest as recent works (e.g. \citealt{kondapally2022,kondapally2025}) found evidence for some LERGs residing in lower-mass, star-forming systems that may have a different fuelling mechanism. Given our significantly larger spectroscopic sample ($\sim70\times$ larger than \citetalias{arnaudova2025}), we are able to examine the distribution of accretion rates and related properties in much greater detail.

To do this, we estimate the log Eddington-scaled accretion rate ($\lambda_{\mathrm{Edd}}$) following \citetalias{arnaudova2025} by assuming that the total energetic output of the black hole includes both the bolometric radiative luminosity ($L_{\mathrm{bol}}$) and the jet mechanical luminosity ($L_{\mathrm{mech}}$), such that $\lambda_{\mathrm{Edd}} = \log_{10}[(L_{\mathrm{bol}} + L_{\mathrm{mech}})/L_{\mathrm{Edd}}]$. Here $L_{\mathrm{Edd}} = 1.3\times10^{31} (M_{\mathrm{BH}}/M_{\odot}$)\,W is the Eddington luminosity, where $M_{\mathrm{BH}}$ is the black hole mass, which we estimate using the stellar mass–black hole mass relation from \citet{haring2004}. The bolometric luminosity is obtained from the observed [\textsc{Oiii}] luminosity using $L_{\mathrm{bol}}~=~3500\,L_{\mathrm{[OIII]}}$ \citep{heckman2004}, where a 3$\sigma$ upper limit is used in cases where [\textsc{Oiii}]$\lambda$5007 is not significantly detected. The jet mechanical power is estimated as $L_{\mathrm{mech}} = 7.3 \times 10^{36} (L_{1.4\mathrm{GHz}}/10^{24}\,\mathrm{W\,Hz}^{-1})^{0.7}$W \citep{cavagnolo2010}, where the 144\,MHz luminosity is converted to 1.4\,GHz assuming a spectral index of $\alpha=-0.7$. We note that each of these scaling relations is characterised by substantial intrinsic scatter, and that the use of [\textsc{O\,iii}] as a proxy for bolometric luminosity may be less reliable for LERGs (e.g. \citealt{heckman2004}) and may differ in low-mass systems (e.g. \citealt{moran2014}). Nevertheless, this framework is widely adopted in the literature (e.g. \citealt{best2012on}; \citealt{mingo2014}; \citealt{Chilufya2024}; \citetalias{arnaudova2025}) to compare LERGs and HERGs, and our sources predominantly reside in massive host galaxies ($M>10^{10}M_{\odot}$).

The resulting $\lambda_{\rm Edd}$ distributions for LERGs and HERGs classified with $>90$ per cent confidence are presented in Figure \ref{fig:lambda_edd}. We additionally show the distributions for the separate LERG classification methods, namely the BPT/$\mathcal{M}$Ex diagnostics and the EW$_{\rm [OIII]}$-based classifications with \textsc{RX\_warning\_frac}<0.10 and 0.10<\textsc{RX\_warning\_frac}<0.50. We note that the BPT/$\mathcal{M}$Ex LERGs approximately represent true measurements of the accretion-rate distributions based on significant [\textsc{Oiii}] emission, whereas the EW$_{\rm [OIII]}$-selected LERGs should be treated as upper limits. In addition, the distribution of the full sample of sources identified as having a radio excess at >90 per cent confidence is presented. This sample is primarily composed of securely classified HERGs and LERGs, but also includes 7,959 additional objects (denoted RX\_Unc) for which the available diagnostics do not allow a confident distinction between the two classes. We find that LERGs display a strong peak at low accretion rates (centred at $\sim$1 per cent of the Eddington limit), while HERGs are predominantly associated with higher $\lambda_{\rm Edd}$ values, as expected (e.g. \citealt{best2012on}; \citealt{mingo2014}; \citetalias{arnaudova2025}). The radio-excess population follows a similar distribution to LERGs at low accretion rates, reflecting their greater numbers, but exhibits a broad tail toward higher $\lambda_{\rm Edd}$ where HERGs become more common, forming a continuous distribution between the two canonical accretion modes. Interestingly, the LERG distribution also shows a non-negligible high–$\lambda_{\rm Edd}$ tail that extends into the HERG regime, which is associated with the BPT/$\mathcal{M}$Ex LERGs and therefore reflects direct accretion-rate measurements rather than upper limits. Although these high-accreting LERGs ($\lambda_{\rm Edd}>-1.5$, corresponding to $\sim 3$ per cent of the Eddington luminosity) represent only a small fraction of the overall LERG population ($\sim$ 9 per cent), their relative abundance is roughly comparable to that of HERGs at similar accretion rates, as reflected in the fraction of the total radio-excess sources in the upper panel\footnote{To verify that the high-accretion LERGs are not misclassified, we examined the sample at a 99 per cent confidence threshold: 9 per cent remain high-accreting. Lowering the threshold (e.g. to 70 or 50 per cent) increases the overlap with HERGs (as observed in \citetalias{arnaudova2025}) and inflates the apparent high–$\lambda_{\rm Edd}$ LERG population, mainly by including otherwise uncertain sources. Adopting a $>90$ per cent threshold therefore minimises contamination and ensures the high-accretion tail represents a robust sub-population.}. We also note that this behaviour is largely driven by the [\textsc{O iii}] or bolometric luminosity, as the same trends are observed when computing the Eddington ratio without including $L_{\rm mech}$.

As noted in Section~\ref{sec:desi_selection}, it is important to account for DESI’s targeting strategy when interpreting these distributions. To assess this, in Appendix~\ref{appendix:desi} we separate the Eddington-scaled distributions into BGS and LRG subsamples. We find that the high--$\lambda_{\rm Edd}$ LERG tail is more prominent in the BGS population, suggesting that these high-accreting LERGs may preferentially reside in systems with more ongoing star formation. We also consider the impact of populations excluded from the main analysis. Type~1 AGN are excluded because our diagnostics are primarily designed for narrow-line systems. The broad-line emission complicates the radio-excess classification, while the strong AGN continuum may bias the SED-derived stellar masses used in the Eddington-scaled accretion-rate estimates. Nevertheless, if all Type~1 AGN were assumed to be radio excess and their stellar masses treated as reliable, they would only strengthen the high-accretion tail of the HERG distribution, as all Type~1 AGN in our sample have $\lambda_{\rm Edd}>-1.5$ (assuming the same calibrations apply). Similarly, as discussed in Appendix~\ref{appendix:unc}, including the large Unc population would primarily increase the sample size of the existing populations without qualitatively changing the main conclusions of this work.

While the aforementioned DESI selection effects may influence the observed $\lambda_{\rm Edd}$ distributions, the large DESI spectroscopic sample nevertheless provides an important opportunity to investigate the properties of the classified radio-source populations, which we present in the following sections.

\subsection{\texorpdfstring{Composite spectra as a function of $\lambda_{\mathrm Edd}$}{Composite spectra as a function of lambdaEdd}}\label{sec:stacks}

\begin{figure*}
    \centering
    \includegraphics[width=1\textwidth]{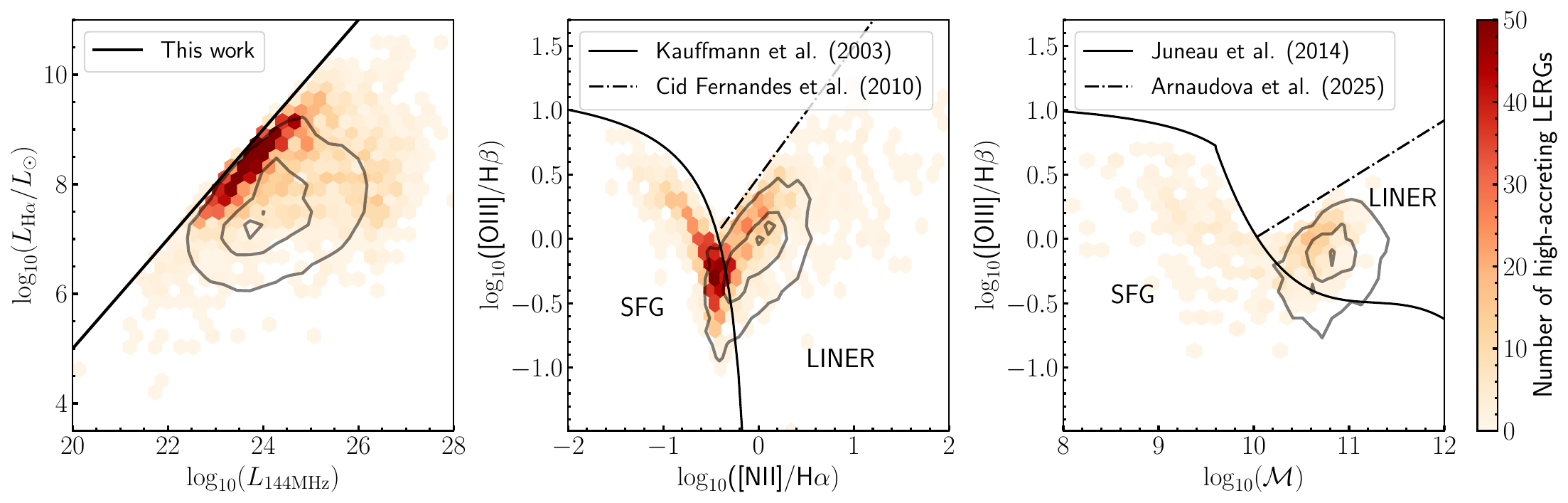}
    \caption{The radio excess (left), BPT (middle) and $\mathcal{M}$Ex diagnostic diagrams (right) for the high-accreting LERGs ($\lambda_{\rm{Edd}}$>-1.5), as denoted by the colour bar. The grey contours encompass 10, 50 and 90 per cent of the low-accreting LERGs. The demarcation lines are the ones used in the classification scheme discussed in section \ref{sec:classifications}, and are denoted by the legend in each panel. For reference, the total number of high-accreting LERGs is 2,611, while the total number of low-accreting LERGs is 25,688. The remaining 6,911 LERGs do not have reliable stellar mass estimates and therefore lack an accretion rate measurement.}
    \label{fig:diagn_lambda}
\end{figure*}

To investigate how the spectral properties of the radio–loud population vary with accretion rate, we use \texttt{\textsc{SpecStacker}}\footnote{\href{https://github.com/m-arnaudova/SpecStacker}{https://github.com/m-arnaudova/SpecStacker}} to construct median-stacked spectra in bins of $\lambda_{\rm Edd}$. To summarise, this stacking code shifts the spectra to the rest frame and resamples them onto a common wavelength grid. Next, it normalises the spectra at the reddest wavelength range where all spectra contribute, and takes the median value at each wavelength to create the stacked spectrum. To estimate uncertainties, \texttt{\textsc{SpecStacker}} uses bootstrapping, along with an additional set of simulations that propagate errors introduced by the redshifting and normalisation steps (see \citealt{arnaudova2024a} for further details). The results of this procedure are shown in Figure \ref{fig:lambda_edd_stacks} for LERGs (top panel) and HERGs (second panel) classified with >90 per cent confidence, radio-excess objects that we cannot confidently classify into a LERG or a HERG (or radio-excess unclassified; third panel), and the total radio excess population (bottom panel) binned in increments of $\Delta \lambda_{\rm Edd} = 0.5$ dex\footnote{We note that due to the broad redshift range considered ($0 < z < 1$), not all spectra contribute to every wavelength bin, resulting in inhomogeneous uncertainties across the stacked spectrum, particularly for the radio-excess unclassified sources.}. To further investigate our results, we applied \texttt{\textsc{WL-SLAYER}} in a setup similar to that described in section \ref{sec:fitting} to measure emission lines and place them on the BPT diagram (see Figure \ref{fig:BPT_lambda_edd}). We consider only fits with an acceptable continuum model ($\chi^{2}_{\nu} < 3$), in order to prevent the emission-line modelling from fitting continuum artefacts\footnote{Stacked spectra, which have higher SNR and contain a mixture of stellar populations, are challenging to model accurately using the BC03 stellar population library.}, and require a $3\sigma$ detection in all BPT emission lines.

In the top panel of Figure \ref{fig:lambda_edd_stacks}, we can see that the LERG stacked spectra at low $\lambda_{\rm Edd}$ show the expected behaviour for this population: the spectra are dominated by an older stellar population, indicated by a strong 4000\AA~ break and prominent Ca H \& K absorption features, with weak or absent emission lines. As $\lambda_{\rm Edd}$ increases, however, emission features begin to emerge, in particular the Balmer lines (H$\delta$, H$\gamma$, H$\beta$, and H$\alpha$), as well as [\textsc{Oii}]$\lambda\lambda3726, 3728$, [\textsc{Oiii}]$\lambda\lambda4959,5007$, [\textsc{Nii}]$\lambda\lambda$6548,6583, and the [\textsc{Sii}]$\lambda\lambda6717,6731$ doublet. The relative strengths of the Balmer and forbidden lines indicate that these galaxies remain in a low-ionisation state, although their emission line ratios gradually evolve from the LINER region toward the SFG region in the BPT diagram\footnote{We note that the same features persist when restricting the analysis to the $0.2 < z < 0.4$ range, suggesting that cosmic evolution does not significantly affect these results.}, as seen in Figure \ref{fig:BPT_lambda_edd}. 
The continuum shape also changes with increasing accretion rate, becoming progressively bluer, accompanied by a weakening of the Ca H \& K absorption, consistent with a growing contribution from a younger stellar population (cf. \citealt{kondapally2025}). 

While a similar trend is seen in the HERG spectra, contrary to the LERGs, the forbidden lines appear much stronger than the Balmer lines (even in the lowest accretion rate stacks, the [\textsc{Oiii}]$\lambda5007$ and [\textsc{Nii}]$\lambda6583$ emission line can be seen, while H$\beta$ and H$\alpha$ are absent), and are located as expected into the Seyfert region in the BPT (see Figure \ref{fig:BPT_lambda_edd}). There are also a number of higher ionisation lines such as [Ne \textsc{v}] and [Ne \textsc{iii}] that are very pronounced in the HERGs, while absent or weak for the LERG stacks for the same accretion rate bin. Similar results were observed when comparing the stacked spectra of HERGs and star-forming LERGs in \citet{kondapally2025}.

The stacked spectra of the radio-excess unclassified objects (third panel) resemble those of HERGs, showing higher-ionisation lines such as [Ne \textsc{v}], albeit at lower strength. Their emission line ratios place them near the boundary between LERGs and HERGs on the BPT diagram, consistent with their intermediate spectral properties.
Finally, the total radio-excess population (bottom panel) exhibits a clear transition between the LERG- and HERG-dominated regimes, reflecting the accretion rates at which each mode becomes more prevalent. At low accretion rates ($\lambda_{\rm Edd} \lesssim -1.5$), the stacked spectra closely resemble those of the LERGs, where the spectra are continuum dominated. As $\lambda_{\rm Edd}$ increases, the spectra gradually shift toward HERG-like characteristics, where the forbidden lines and in particular higher ionisation lines become more prominent, which is clearly seen in the BPT diagram. 

Overall, these results highlight that, within a given $\lambda_{\rm Edd}$ bin, LERGs and HERGs remain distinct, while the emergence of Balmer lines and bluer continua in high-accreting LERGs points to ongoing star formation, helping to explain their increased presence in the BGS sample.

\subsection{High-accreting LERGs as a function of diagnostics}\label{sec:diag_lambda}

Given the results of the previous section, we examined the location of high-accreting LERGs ($\lambda_{\rm Edd} > -1.5$) in the diagnostics introduced in Section \ref{sec:classifications}, in addition to the BPT results from the stacked spectra shown in Figure \ref{fig:BPT_lambda_edd}, to characterise their properties in more detail. For the radio-excess diagnostic, we consider the full sample of high-accreting LERGs, whereas for the BPT and $\mathcal{M}$Ex diagrams we include only sources classified at $>90$ per cent confidence in the respective diagnostics. The [\textsc{O iii}] EW method does not yield a significant population of high-accreting LERGs and is therefore excluded from this analysis.

The results are presented in Figure \ref{fig:diagn_lambda}, where we can see in the left panel that the majority of high-accreting LERGs, shown as the colour scale, are concentrated near the radio-excess demarcation line, suggesting that their radio emission may be modestly enhanced relative to their H$\alpha$ luminosity, potentially due to low-power AGN activity or contributions from intense star-formation driven shocks. By contrast, low-accreting LERGs ($\lambda_{\rm Edd} < -1.5$), represented by grey contours, populate a broader region of the radio-excess plane, with sources both near the demarcation line and at higher radio powers for a given H$\alpha$ luminosity.
In the BPT diagram (middle panel), high-accreting LERGs predominantly occupy the BPT\_SFG region, while in the $\mathcal{M}$Ex diagram (right panel) they are distributed between the $\mathcal{M}$Ex\_SFG and $\mathcal{M}$Ex\_LINER regions. This mixed behaviour may be influenced by redshift-dependent effects, since the $\mathcal{M}$Ex diagram is used primarily at $z>0.5$, or by optical selection biases (e.g. LRGs and BGS subsamples). Low-accreting LERGs, in contrast, are mostly found in the BPT\_LINER and $\mathcal{M}$Ex\_LINER regions, suggesting that they are associated with different ionisation properties.

Taken together, these results indicate that high-accreting LERGs may either be star-forming galaxies with a modest radio excess due to shocks or recent star formation, or genuine LERGs at lower radio powers residing in more actively star-forming hosts. In either scenario, star-forming processes could contribute to the [\textsc{O iii}] emission, potentially leading to an over-estimation of $L_{\rm [OIII]}$ associated with the AGN activity and thus $\lambda_{\rm Edd}$. Furthermore, as previously discussed, the conversion from $L_{\rm [OIII]}$ to $L_{\rm bol}$ is subject to significant intrinsic scatter and may be less reliable for LERGs. If this calibration does not hold in this regime, the inferred bolometric luminosities and consequently the Eddington ratios may not be representative of the true accretion power. Future work using independent constraints on the accretion power, for example from X-ray measurements, would therefore provide a valuable cross-check. We cannot also rule out the possibility that, if these systems are indeed LERGs, their fuelling mechanisms may differ from the standard hot-gas accretion via advection-dominated flows (ADAFs; \citealt{narayan1994, narayan1995}) observed in typical low-accreting LERGs. These results will be explored further by Kondapally et al. (\textit{in prep.}).

\section{Conclusions}\label{sec:summary}

In this work, we have used the spectroscopic information from the first data release of the Dark Energy Spectroscopic Instrument (DESI) survey to probabilistically classify the radio source population observed by the second data release of the LOFAR Two-metre Sky Survey (LoTSS). Building upon the classification methods by \citetalias{drake2024lofar} and \citetalias{arnaudova2025}, we have used a combination of radio excess, the BPT diagnostic, and a modified MEx ($\mathcal{M}$Ex) diagnostic, alongside Monte Carlo methods to estimate the probability of a source being a star-forming galaxy (SFG), a radio-quiet AGN (RQ AGN), or a low- or high-excitation radio galaxy (LERG or HERG). In addition, we used the [\textsc{Oiii}]$\lambda5007$ equivalent width to recover LERGs associated with low-significance emission lines, significantly increasing the sample size and enabling a more complete recovery of the LERG population. Applying a 90 per cent probability threshold to this classification framework, we identify 68,820 SFGs, 32,288 RQ AGN, 35,210 LERGs and 3,085 HERGs. 

Using this high-confidence DESI-LOFAR sample, we:
\begin{itemize}
    \item Compared our classifications with \citetalias{drake2024lofar}, which are based on SDSS spectroscopy, where we found comparable results when considering the sources classified by both works. However, some discrepancies arise when including the unclassified fraction, resulting from our use of aperture corrections, additional diagnostics and the more rigorous treatment of the Monte Carlo realisations.

    \item Assessed the impact of DESI selection effects by comparing the properties of the numerically dominant samples from DESI's Bright Galaxy Survey (BGS) and Luminous Red Galaxy (LRG) surveys, showing that selection biases affect individual class distributions, and thus must be taken into account.

    \item Examined the Eddington-scaled accretion rate ($\lambda_{\rm Edd}$) distributions up to $z=1$, confirming very distinct accretion rate distributions for the two classes of objects where LERGs typically accrete below 1 per cent of Eddington and HERGs above it with higher statistical significance than before.

    \item Uncovered a high-accreting population of LERGs that appear to be hosted by more strongly star-forming systems, as shown by the stacked spectra and their location on the diagnostic diagrams. These high-accreting LERGs appear similar to the star-forming LERG population identified by \citet{kondapally2022, kondapally2025}, despite differences in selection and classification methods. We outline several possible explanations for their nature, including contamination of the [\textsc{O iii}] emission by star-forming processes, limitations in the $L_{\rm [OIII]}$–$L_{\rm bol}$ conversion in this regime, or a distinct fuelling mechanism compared to typical low-accreting LERGs. Future multi-wavelength observations, particularly in the X-ray, will be essential to disentangle these scenarios.

\end{itemize}

In addition to the findings reported in this paper, our catalogue will be invaluable for statistical investigations of other key aspects of galaxy evolution. For example, we are currently undertaking a study of the properties of AGN-driven outflows for the different classifications presented here (Holden et al., \textit{in prep.}), which is now possible thanks to the large sample size provided by DESI. Building on this work, the upcoming WEAVE-LOFAR survey (\citealt{smith2016}) will directly target LoTSS sources, selecting solely on radio flux density and thus offering a more complete and unbiased view of the radio source population. By reducing the optical selection biases inherent in the current DESI-based sample, WEAVE-LOFAR will allow robust characterisation of radio sources’ physical properties, accretion modes, and host galaxies across a wide range of environments. This will open the door to more detailed studies of high-accreting LERGs, and more broadly to a deeper understanding of the interplay between star formation and AGN activity and their impact on galaxy evolution.

\section*{Acknowledgments}
The authors sincerely thank the anonymous reviewer for a report which significantly improved the quality of this paper.
The authors would also like to thank Lorenzo R. Tagliapietra for his contribution to the comparison of the H$\alpha$ and W3-based radio-excess diagnostics.

MIA and PNB acknowledge support from the UK Science
and Technology Facilities Council (STFC) under grant ST/Y000951/1.
DJBS and LRH acknowledge support from STFC under grant ST/Y001028/1.
DJBS acknowledges support from STFC grant ST/V000624/1.
DJBS and SD acknowledge support from the Leverhulme Trust via Research Project Grant RPG-2025-078.
PNB is grateful for support from the UK STFC via grant ST/V000594/1.
RK acknowledges support from the Leverhulme Trust through a Leverhulme Trust Early Career Fellowship.
KJD acknowledges support from STFC through an Ernest Rutherford Fellowship (grant number ST/W003120/1).
MJH thanks the UK STFC for support via grant ST/V000624/1 and ST/Y001249/1.
SS and DJBS acknowledge support from the UK STFC via the grant ST/X508408/1.
CLH acknowledges support from STFC through grant ST/Y000951/1.

LOFAR is the Low Frequency Array designed and constructed by ASTRON. It has observing, data processing, and data storage facilities in several countries, which are owned by various parties (each with their own funding sources), and which are collectively operated by the LOFAR ERIC under a joint scientific policy. The LOFAR resources have benefited from the following recent major funding sources: CNRS-INSU, Observatoire de Paris and Université d'Orléans, France; BMBF, MIWF-NRW, MPG, Germany; Science Foundation Ireland (SFI), Department of Business, Enterprise and Innovation (DBEI), Ireland; NWO, The Netherlands; The Science and Technology Facilities Council, UK; Ministry of Science and Higher Education, Poland; The Istituto Nazionale di Astrofisica (INAF), Italy.
This research made use of the Dutch national e-infrastructure with support of the SURF Cooperative (e-infra 180169) and the LOFAR e-infra group. The Jülich LOFAR Long Term Archive and the German LOFAR network are both coordinated and operated by the Jülich Supercomputing Centre (JSC), and computing resources on the supercomputer JUWELS at JSC were provided by the Gauss Centre for Supercomputing e.V. (grant CHTB00) through the John von Neumann Institute for Computing (NIC).
This research made use of the University of Hertfordshire high-performance computing facility and the LOFAR-UK computing facility located at the University of Hertfordshire and supported by STFC [ST/P000096/1], and of the Italian LOFAR IT computing infrastructure supported and operated by INAF, and by the Physics Department of Turin university (under an agreement with Consorzio Interuniversitario per la Fisica Spaziale) at the C3S Supercomputing Centre, Italy.
This research is part of the project LOFAR Data Valorization (LDV) [project numbers 2020.031, 2022.033, and 2024.047] of the research programme Computing Time on National Computer Facilities using SPIDER that is (co-)funded by the Dutch Research Council (NWO), hosted by SURF through the call for proposals of Computing Time on National Computer Facilities. 

\section*{Data Availability}

The radio and spectroscopic data used in this work are publicly available from the LOFAR Surveys (\href{https://www.lofar-surveys.org}{https://www.lofar-surveys.org}) and the Dark Energy Spectroscopic Instrument (\href{https://data.desi.lbl.gov}{https://data.desi.lbl.gov}), respectively. The output catalogue, including the emission-line measurements and the classification probabilities, as well as the stacked spectra of the radio-excess population will be made available as part of this paper at \href{https://lofar-surveys.org/dr2_release.html}{https://www.lofar-surveys.org/dr2\_release.html}.



\bibliographystyle{mnras}
\bibliography{ref} 



\appendix

\begin{figure}
    \centering
    \includegraphics[width=1\columnwidth]{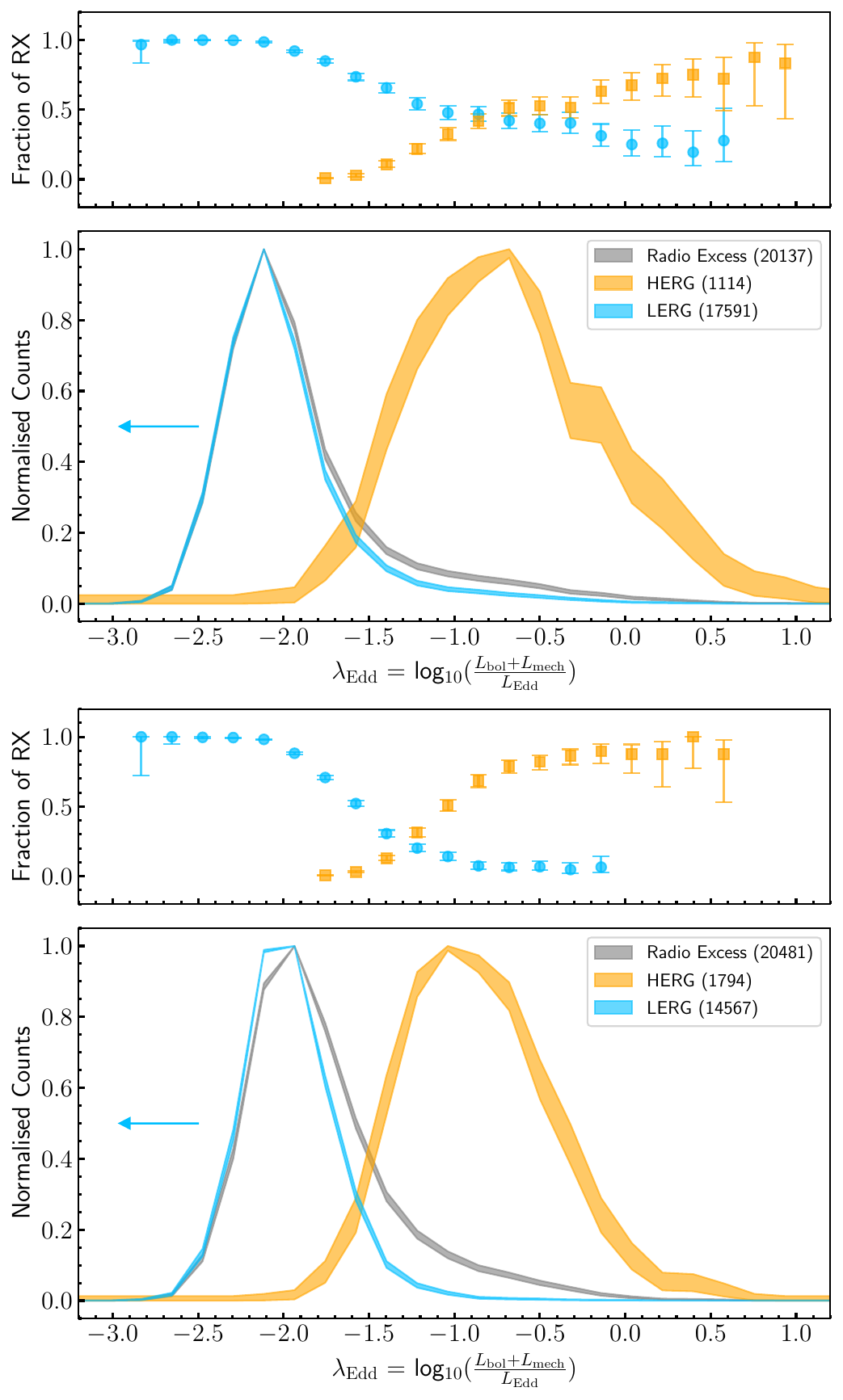}
    \caption{The Eddington-scaled accretion rate distributions for LERGs (blue), HERGs (orange), and all radio-excess sources (grey), separated into the subsamples belonging to the DESI BGS (top) and LRG (bottom) survey selections. The arrows indicate that the LERG accretion rates should be considered upper limits due to the $3\sigma$ upper limit on the [\textsc{Oiii}] emission, while the shaded regions indicate binomial uncertainties. The upper panel shows the fraction of radio-excess sources classified as LERGs (blue) and HERGs (orange), with error bars representing binomial uncertainties. Only bins containing at least five sources are included.}
    \label{fig:desi_selection_acc_rates}
\end{figure}

\section{Impact of DESI selection}\label{appendix:desi}
To evaluate the impact of DESI’s targeting strategy on the Eddington-scaled accretion rates, we separate our radio-loud sample into the two numerically dominant surveys—BGS and LRGs—and compute the accretion-rate distributions for LERGs, HERGs, and radio-excess (RX) sources following the procedure outlined in Section \ref{sec:accretion_rates}. The results are shown in Figure \ref{fig:desi_selection_acc_rates},
where we show the Eddington-scaled accretion-rate distributions (with source counts listed in the legends), alongside the fraction of RX sources classified as LERGs or HERGs in each subsample.

In both the BGS and LRG samples, we recover the characteristic dual peaks associated with LERGs and HERGs. However, in the BGS sample the high-$\lambda_{\rm Edd}$ tail of the LERG distribution is much more pronounced. This suggests that the high-accreting LERGs may reside in more star-forming systems, which are galaxy types that the LRG selection is designed to exclude. We explore this further in sections~\ref{sec:stacks} and \ref{sec:diag_lambda}.

\section{Impact of unclassified sources}\label{appendix:unc}

\begin{figure}
    \centering
    \includegraphics[width=1\columnwidth]{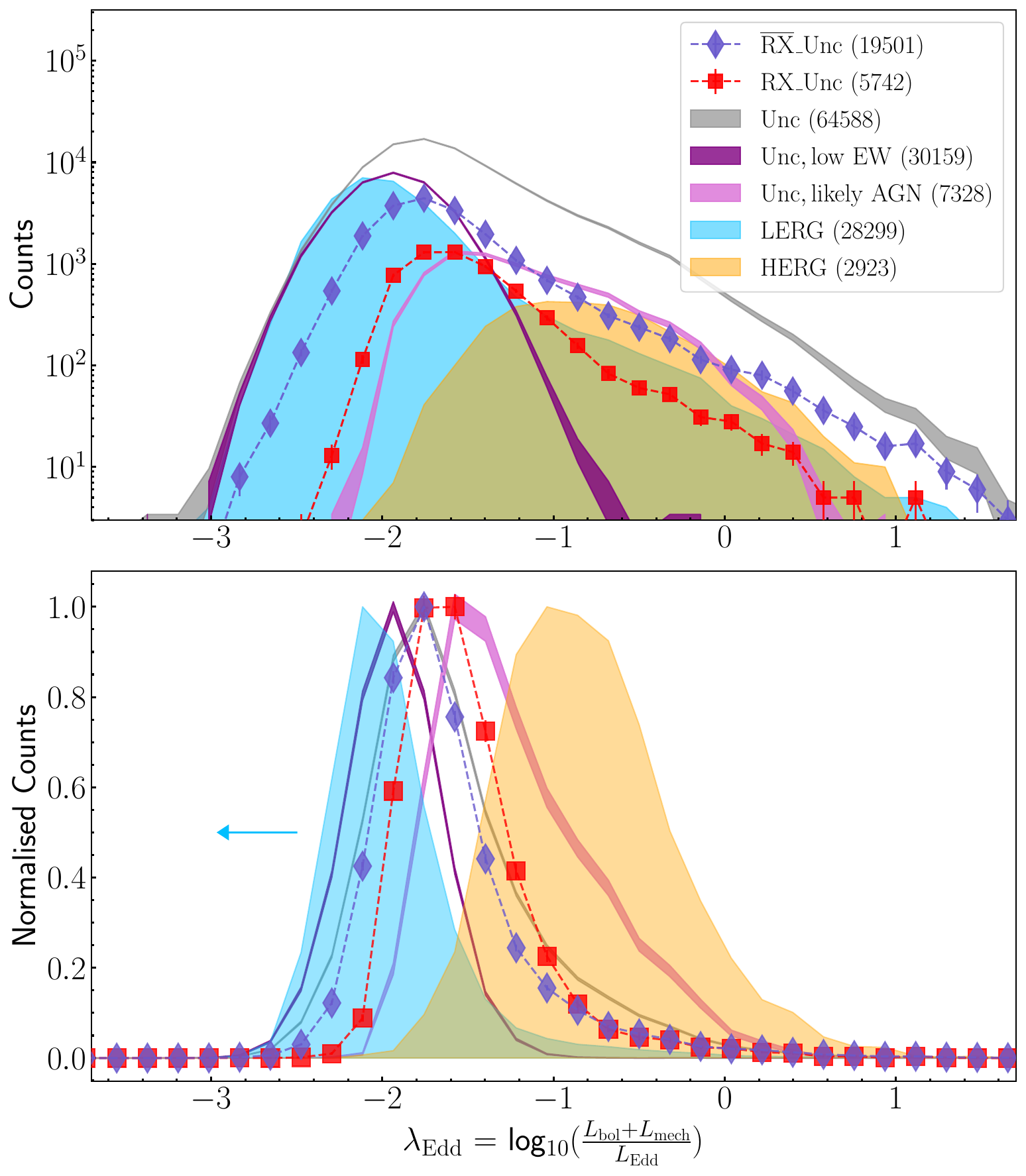}
    \caption{The Eddington-scaled accretion-rate total-count (upper panel) and normalised (lower panel) distributions for radio-excess sources that cannot be divided into LERGs and HERGs (uncertain radio-excess sources, RX\_Unc; red) at 90 per cent confidence, non-radio-excess sources that cannot be confidently classified as either SFGs or RQ AGN (uncertain non-radio-excess sources, $\overline{\rm{RX}}$\_Unc; dark blue), and the remaining unclassified sources (Unc; grey), further split into low-EW (purple) and ``likely'' AGN (magenta) subsets. The distributions of LERGs (light blue) and HERGs (orange) are also shown for reference. The corresponding Poisson (upper panel) and binomial (lower panel) uncertainties are shown as error bars and shaded regions, respectively. The number of sources in each class is given in the legend in the upper panel; these numbers differ from Table \ref{tab:target_classification} as not all sources have reliable stellar mass estimates. }
    \label{fig:acc_rates_unc}
\end{figure}

As discussed in section~\ref{sec:classifications}, we have a large fraction of unclassified sources which lie close to the boundaries of our diagnostics and therefore cannot be confidently assigned to a given class at the adopted \(>90\) per cent reliability threshold without additional information. Nevertheless, it is informative to investigate the impact of excluding these systems from the final analysis. In Figure~\ref{fig:acc_rates_unc}, we present the accretion-rate distributions of: (i) radio-excess sources that cannot be robustly separated into LERGs and HERGs (RX\_Unc), (ii) non-radio-excess sources that cannot be confidently classified as either SFGs or RQ AGN ($\overline{\rm{RX}}$\_Unc), and (iii) the remaining unclassified sources (Unc), for which even the presence of a radio excess cannot be determined at the 90 per cent confidence level. The figure shows both the total-count and normalised distributions, allowing comparison of both the absolute numbers and relative distribution shapes of the different populations. We note that 42,338 of the 251,413 sources in our sample (corresponding to 19, 28, 21, 20, and 5 per cent of the Unc, RX\_Unc, $\overline{\rm{RX}}$\_Unc, LERG, and HERG populations, respectively) do not have reliable stellar-mass estimates from \textsc{CIGALE} and are therefore excluded from the accretion-rate analysis presented in this work. These sources do not appear to occupy a distinct region of parameter space in redshift, $L_{\rm144\,MHz}$, or [\textsc{Oiii}]$\lambda$5007 luminosity.

We find that the overall Unc distribution is more similar to that of the $\overline{\rm{RX}}$\_Unc population, suggesting that the fully unclassified sources are likely dominated by systems without a strong radio excess and therefore predominantly by SFG-like galaxies. However, if this population were associated with radio-excess systems, we can investigate possible subsets within the Unc population. One such subset consists of systems with weak emission lines, analogous to the weak-lined LERGs discussed previously, but without requiring a radio excess. Specifically, we select sources with EW$_{\rm [OIII]}<3$\AA~ in more than 90 per cent of the Monte Carlo realisations (\textsc{EW\_frac}>0.90). This criterion is satisfied by approximately 50 per cent of the Unc population and the associated Eddington-scaled accretion-rate distribution of this population appears more LERG-like (see Figure~\ref{fig:acc_rates_unc}). In addition, Figure~\ref{fig:demographics1} shows that essentially no SFGs are identified at \(z>0.5\), \(\log_{10}(L_{144\rm MHz}/{\rm W\,Hz^{-1}})\gtrsim24\), and \(\log_{10}(M_{*}/M_{\odot})\gtrsim11\), a parameter space typically associated with radio AGN (e.g. \citealt{sabater2019}). This ``likely'' AGN subset represents approximately $\sim10$ per cent of the Unc population and, as can be seen in Figure~\ref{fig:acc_rates_unc}, its $\lambda_{\rm Edd}$ distribution peaks near the transition between the LERG and HERG populations ($\lambda_{\rm Edd}\sim-1.5$), similar to the RX\_Unc population, but also extends to higher Eddington-scaled accretion rates, suggesting that it comprises a mixture of sources with intermediate properties and more HERG-like systems. The sources near the transition likely represent objects with intermediate properties, as is also apparent from the RX\_Unc stacked spectra presented in Section~\ref{sec:stacks}.

Overall, this analysis suggests that excluding the fully unclassified population is unlikely to qualitatively change the main conclusions of this work. Instead, including these sources would primarily increase the sample size of the existing populations, with most Unc sources appearing consistent with LERG-like systems and a smaller subset occupying the transition region between the LERG and HERG populations or extending into the HERG regime.


\bsp	
\label{lastpage}
\end{document}